\documentclass[times]{stvrauth}

\usepackage[colorlinks,bookmarksopen,bookmarksnumbered,citecolor=red,urlcolor=red]{hyperref}

\usepackage{graphicx}
\usepackage{subfig}
\DeclareMathOperator{\logit}{logit}

\newcommand{\mr}{\mathrm}
\usepackage{amsfonts}
\usepackage{IEEEtrantools}
\usepackage[ruled,vlined]{algorithm2e}
\usepackage{footnote}

\begin{document}

\runningheads{M.~Scholz and R.~Torkar}{An empirical study of Linespots}

\title{An empirical study of Linespots: A novel past-fault algorithm}

\author{Maximilian Scholz\affil{1}\corrauth,
Richard Torkar\affil{2,3}}

\address{
\affilnum{1}University of Stuttgart, Cluster of Excellence SimTech - Research Group for Bayesian Statistics, Stuttgart, Germany, \break
\affilnum{2}Chalmers and University of Gothenburg, Dept.~of Computer Science and Engineering, Göteborg, Sweden \break
\affilnum{3}Stellenbosch Institute for Advanced Study (STIAS), Stellenbosch, South Africa
}

\corraddr{maximilian.scholz@simtech.uni-stuttgart.de}

\begin{abstract}
This paper proposes the novel past-faults fault prediction algorithm Linespots, based on the Bugspots algorithm. We analyze the predictive performance and runtime of Linespots compared to Bugspots with an empirical study using the most significant self-built dataset as of now, including high-quality samples for validation. As a novelty in fault prediction, we use Bayesian data analysis and Directed Acyclic Graphs to model the effects.

We found consistent improvements in the predictive performance of Linespots over Bugspots for all seven evaluation metrics.
We conclude that Linespots should be used over Bugspots in all cases where no real-time performance is necessary.
\end{abstract}

\keywords{Linespots, Fault Prediction, Past Faults, Bugspots, Bayesian Data Analysis, Directed Acyclic Graphs}

\maketitle

\section{Introduction}
The field of fault prediction has brought up many different metrics to base predictions on. From simple size-based metrics like lines of code (LOC), complexity metrics like McCabe and Halstead, object-oriented metrics such as coupling and cohesion, all the way to process metrics like code churn, socio-technical networks, and development history~\cite{hallSystematicLiteratureReview2012,radjenovicSoftwareFaultPrediction2013}. In this paper we will focus on one specific category of temporal metrics, i.e., past faults~\cite{hassanTopTenList2005, kimPredictingFaultsCached2007a}. Rahman et al.~\cite{rahmanBugCacheInspectionsHit2011} showed that they perform similar to the more complex BugCache algorithm and are easier to compute. Since then, multiple studies have found past faults to be good predictors for future faults~\cite{dambrosExtensiveComparisonBug2010, dambrosEvaluatingDefectPrediction2012}.

Based on the idea by Rahman et al.~\cite{rahmanBugCacheInspectionsHit2011}, Lewis et al.~\cite{lewisDoesBugPrediction2013} developed the Bugspots algorithm, which added a weight decay for older faults. In this paper, we present Linespots, a novel fault prediction algorithm based on Bugspots. As the name hints, Linespots runs on a line-level granularity instead of files, as is the case with Bugspots. This focus on lines follows from the insights from Kochhar et al.~\cite{kochharPractitionersExpectationsAutomated2016} who showed that developers prefer more granularities in results, and Hata et al.~\cite{hataBugPredictionBased2012}, who showed a trend in better performance when using finer granularities.

In this paper, we present an empirical study of Linespots' predictive performance and runtime compared to Bugspots. Focusing on predictive performance and runtime is due to them being considered the main benefits of Bugspots. In order to conduct such a comparative analysis, we have spent considerable effort in collecting representative data.

While there are pre-built datasets to benchmark fault prediction metrics and algorithms (e.g., PROMISE \cite{shirabadPROMISERepositorySoftware2005}), many of them do not provide the necessary information for Linespots. Additionally, more involved datasets, like defects4j~\cite{justDefects4JDatabaseExisting2014}, lack in sample size. To avoid such limitations, this study uses an open-source dataset and self-built validation data. While self-built validation data can be unreliable~\cite{dambrosExtensiveComparisonBug2010}, we used high-quality samples to vet the collected validation data.

As a novelty in the field of fault prediction, we use Bayesian data analysis to model the effects of different algorithms and data quality and find improvements in critical areas for Linespots over Bugspots. When applying Bayesian data analysis, we follow the recommendations of Wasserstein et al.~\cite{wassersteinMovingWorld052019} to abandon the use of $p$-values and point estimates.

The key contributions of this paper are:

\begin{itemize}
    \item The proposal of Linespots, a novel fault prediction algorithm.
    \item Presentation of the results from a comparative study analyzing performance and runtime using Bugspots and Linespots as artifacts.
    \item Use of Bayesian data analysis and Directed Acyclic Graphs (DAGs) to model different algorithms' effects on evaluation metrics.
    \item Validation of a self-built dataset with a high-quality sub-sample.
    \item Finally, to our knowledge, this study uses the largest dataset published, based on the revision count of different projects in the field, as of now~\cite{liEvaluatingSoftwareDefect2019}. 
\end{itemize}

The research questions we pose are:
\begin{itemize}
    \item RQ1: How does the predictive performance of Linespots compare to Bugspots?
    \item RQ2: How does the runtime performance of Linespots compare to Bugspots?
\end{itemize}

In the following background section (Section~\ref{sec:background}), we will present the history and details of Bugspots and Linespots. Section~\ref{sec:method} provides details concerning study design, including the research questions, our process for gathering and preparing the dataset, and, finally, a high-level explanation of our analysis.

We present our results in Section~\ref{sec:results} and explain threats to the study validity in Section~\ref{sec:threats-to-validity}.
Following that, we discuss the implications, of said results, in Section~\ref{sec:discussion}. In Related Work (Section~\ref{sec:related-work}), we explain the connection to other work. Finally, we , before presenting a summary and our conclusions, together with suggestions for future work in Section~\ref{sec:conclusion}.

We offer a reference implementation of Linespots together with the evaluation code~\cite{LinespotsReferenceImplementation}, a replication package~\cite{scholzLinespotsReproducabilityDocker2020}, and a detailed analysis report~\cite{scholzLinespotsAnalysisRepository2019}, as supplements to this paper.

\section{Background}
\label{sec:background}
Fault prediction is the process of using statistical techniques to predict where faults might occur in the code of a project.
The process of prediction can classify parts of the code as faulty or not faulty, can flag fault-inducing modifications to the projects or rank code elements by their likelihood of containing faults.
Common approaches for fault prediction are based on static code metrics like size and complexity and object-oriented metrics like coupling, cohesion and inheritance.
More recently, process metrics that include development process information, are being developed. This includes the past faults class that Bugspots and Linespots are a part of, as well as concepts like code churn and socio-technical networks \cite{hallSystematicLiteratureReview2012}.
Before going into details about the Bugspots and Linespots algorithms, we want to present our use of the terms metric, model, and algorithm to avoid confusion. As the terms have not been used with the same meaning consistently throughout past works in fault prediction and localization, these are the definitions for how we use the terms:

\begin{itemize}
    \item Algorithm: Algorithms are procedures that compute fault prediction metrics. In this study, Linespots and Bugspots refer to the algorithms as outlined in Algorithms~\ref{alg:bugspots}--\ref{alg:linespots} and not to the resulting fault prediction metrics.
    \item Fault prediction metric: These are metrics like LOC, complexity, class size, and past-faults, that can be used to predict faults. Linespots and Bugspots both output a version of the past-faults fault prediction metric.
    \item Evaluation metric: Evaluation metrics are the metrics we calculate to compare different fault prediction metrics' performance. The evaluation metrics we use are described in Section~\ref{sec:study-design}.
    \item Model: These are the statistical models we built to analyze the effects of the fault prediction metrics on the evaluation metrics, as described in Section~\ref{sec:analysis}.
    \item Fault prediction model: A combination of multiple fault prediction metrics.
    
\end{itemize}

\subsection{Bugspots}
The idea of using past faults to predict future faults was pioneered by Hassan et al.~\cite{hassanTopTenList2005}. A comparison of past faults metrics with other fault prediction metrics by D'Ambros et al.~\cite{dambrosEvaluatingDefectPrediction2012} showed that past faults metrics could perform well and Rahman et al.~\cite{rahmanBugCacheInspectionsHit2011} found that ranking files by the number of past faults could perform as well as their more complex \textit{FixCache}.

To the basic idea of ranking files by the number of past faults, Lewis et al.~\cite{lewisDoesBugPrediction2013} added a weight decay so that more recent faults would impact the ranking more than faults further in the past. This was done to allow files that contained faults in the past but were fixed to move down in rank over time. The algorithm by Lewis et al.~\cite{lewisDoesBugPrediction2013} has since been used in multiple other studies~\cite{dambrosEvaluatingDefectPrediction2012, wangVersionHistorySimilar2014, youmImprovedBugLocalization2017, zouEmpiricalStudyFault2019}.
Finally, Grigorik~\cite{grigorikImplementationSimpleBug2019} wrote an implementation and called it Bugspots, which is the namesake of our newly proposed algorithm, Linespots.
A pseudo-code representation of Bugspots is shown in Algorithm~1.
The exponential weighting function was chosen to focus most of the weight on the last 6 months of development. This would have to be modified depending on the project. In practice we found that most applications would instead just use the last 500 commits, based on the default from the Bugspots implementation. We found that different weighting functions had little impact on the results in this case so we stuck with the default one to add less variance.

\begin{algorithm}
\SetKwInOut{Input}{input}
\SetKwInOut{Output}{output}
\SetKwFunction{InducesFix}{InducesFix}
\SetKwFunction{Age}{Age}
\SetKwFunction{SortByScore}{SortByScore}
\SetKwData{Commit}{commit}
\SetKwData{File}{file}
\SetKwData{F}{$F$}
\SetAlgoLined
\Input{A list of commits $C$ ordered by age}
\Output{A ranked list of files}
\BlankLine
$\F = [\ ]$\\
\ForEach{\Commit in $C$}{
    \If{\InducesFix{\Commit}}{
        \ForEach{\File in \Commit}{
            \If{\File not in \F}{
                $\F[\File]=0$
            }
            $\F[\File]=1/(1+\exp((-12*\Age{\Commit})+12))$
        }
    }
}
\KwRet{\SortByScore{\F}}
\caption{Bugspots, simplified}
\label{alg:bugspots}
\end{algorithm}

\subsection{Linespots}
As simple and promising Bugspots seems, the limitation of reporting on the file level granularity was an immediate and obvious drawback. From our experience there are usually only a few faulty lines in an entire file, so that there is a lot of false positives bundled in with the few faulty lines. This gave the basic idea for Linespots: reporting on line level granularity. This move is supported by the findings of Hata et al.~\cite{hataBugPredictionBased2012} who showed that finer granularity gave better prediction results.
As tracking line changes is not as straight forward as file changes, we want to discuss some of the decisions we made when developing the algorithm in this section.

As we work with the Git commit history of the projects, we get all of the information we use from there.
Each Git commit contains some metadata and the file diffs, which describe the changes that the commit applies to the files. Each file diff is further split into hunks, sections of changed code with some padding before and after, for context.
There are different algorithms to calculate the diffs and hunks, the Myers algorithm~\cite{myersAnONDDifference1986} which is used by default is just one of them. We chose to use the Histogram diff algorithm by Nugroho et al.~\cite{nugrohoHowDifferentAre2020} who found that it better reflects the intention of developers for code changes and recommend it over the Myers algorithm for repository mining. However we found that the two algorithms lead to very similar results when briefly comparing them.

Each line in a project starts with a score of 0 that is increased each time a line is part of a fix inducing commit. This is not trivial as git does not support line modifications and instead removes and adds lines.
Starting with removed lines, Linespots simply removes the corresponding scores from the file's score list.
Added lines could either be modified or truly new lines. We do not differentiate between the two and calculate the score of added lines by using the mean hunk score and applying the score increased based on the weighting function to it.
Finally we assume that unmodified lines that are close to a fix inducing modification share the risk of being faulty and thus increase their score based on the weighting function as well. We use the hunk offered by git to define closeness out of convenience.

A pseudo-code representation of Linespots is shown in Algorithm \ref{alg:linespots}.
An open-source reference implementation~\cite{LinespotsReferenceImplementation} is available for those interested in implementation details.

\begin{algorithm}
\SetKwInOut{Input}{input}
\SetKwInOut{Output}{output}
\SetKwFunction{InducesFix}{InducesFix}
\SetKwFunction{Age}{Age}
\SetKwFunction{SortByScore}{SortByScore}
\SetKwFunction{TrackLineChanges}{TrackLineChanges}
\SetKwFunction{Flatten}{Flatten}
\SetKwFunction{Length}{Length}
\SetKwData{Commit}{commit}
\SetKwData{File}{file}
\SetKwData{Hunk}{hunk}
\SetKwData{Line}{Line}
\SetKwData{F}{$F$}
\SetAlgoLined
\Input{A list of commits $C$ ordered by age}
\Output{A ranked list of lines}
\BlankLine
$\F = []$ \\
\ForEach{\Commit in $C$}{
    \ForEach{\File in \Commit}{
        \If{$\F[\File]$ is empty}{
            $\F[\File]=  [0] \times \Length{\File}$
        }
        \ForEach{\Hunk in \File}{
            $\F[\File]=$\TrackLineChanges{$\F[\File]$,\Hunk}
            
            \If{\InducesFix{\Commit}}{
                $\F[\File][\Hunk]= \F[\File][\Hunk] + 1/(1+\exp((-12*\Age{\Commit})+12))$
                
            }
        }
    }
}
\KwRet{\SortByScore{\Flatten{\F}}}
\caption{Linespots, simplified}
\label{alg:linespots}
\end{algorithm}

\section{Method}
\label{sec:method}

\begin{figure}
\centering
   \includegraphics[width=0.95\textwidth]{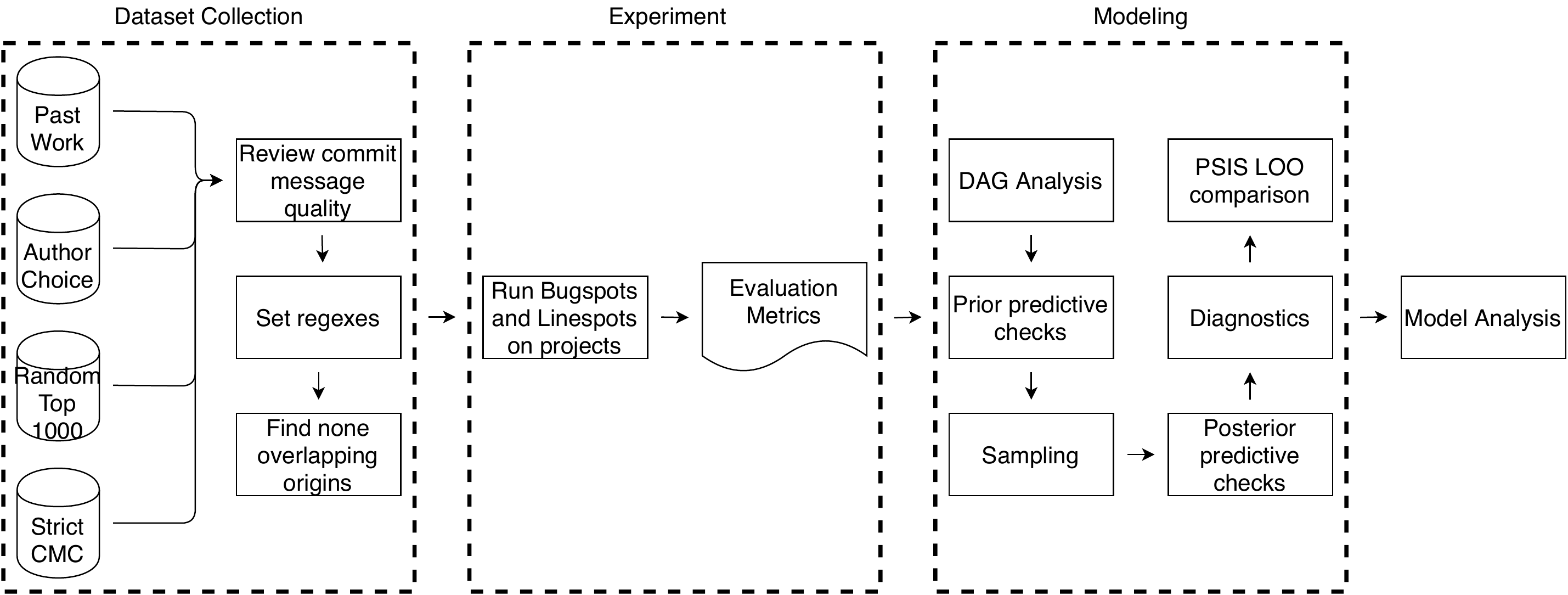}
    \caption{Overview of the study design}
    \label{fig:study-flow}
\end{figure}

This study gathered samples from different projects for both Bugspots and Linespots and compared their predictive and runtime performance.
This section describes the study design, our sampling strategy, and our modeling and analysis approach.
Figure~\ref{fig:study-flow} shows an overview of the complete study.

\subsection{Study design}
\label{sec:study-design}
The goal of this study is to evaluate and compare Bugspots and Linespots concerning predictive performance and runtime.
To focus our efforts, we investigate the following research questions:
\begin{itemize}
    \item RQ1: How does the predictive performance of Linespots compare to Bugspots?
    \item RQ2: How does the runtime performance of Linespots compare to Bugspots?
\end{itemize}

To evaluate the predictive performance, we use common evaluation metrics from the field of fault prediction:

\subsubsection{Area under the receiver operating characteristic curve} (\texttt{AUROC})~\cite{fawcettROCGraphsNotes2004}.
The \texttt{AUROC} is a way to assess the performance of a ranking algorithm by calculating the precision and recall for all possible cut-off points and plotting them against each other. In other words, we plot the precision-recall points from treating only the highest ranked result as faulty up to treating all results as faulty.
An optimal ranking would result in an area of $1$ under the curve, with only faulty elements at the top of the ranking.
While we have not extensively tested this, we assume that it is not valid to compare \texttt{AUROC} values between studies that use different granularities. \texttt{AUROC} values are usually not normalized, so any differences in the artifacts might bias the comparison.

\subsubsection{Area under the cost-effectiveness curve} (\texttt{AUCEC})~\cite{arisholmDataMiningTechniques2007}. The \texttt{AUCEC} is a performance measure that accounts for effort. The cost-effectiveness curve results from going through the result list of a fault prediction algorithm and plotting the proportion of LOC on the $x$-axis and the proportion of found faults on the $y$-axis.
Better performing algorithms will result in a curve with a higher slope.
To compare two curves, the area under the curves can be calculated as a summarization, as a higher sloped curve will also lead to a larger area under the curve.
To improve the reliability when comparing different projects, Arisholm et al.~\cite{arisholmSystematicComprehensiveInvestigation2010} proposed normalizing the AUCEC to the optimal curve achievable with the project as such:

\begin{equation}
\mr{CE}_{\pi} = \frac{\mr{CE}_{\pi}(\mr{model})-\mr{CE}_{\pi}(\mr{baseline})}{\mr{CE}_{\pi}(\mr{optimal})-\mr{CE}_{\pi}(\mr{baseline})}
\end{equation}

\noindent where $\mr{CE}_{\pi}(x)$ is the area under the curve $x$ (baseline, model, or optimal) for a given $\pi$ percentage of LOC\@. The baseline curve represents faults randomly distributed throughout the results such that the cost-effectiveness curve matches $y=x$. The model curve is the one derived from the results of Bugspots and Linespots in this case.

The optimal curve has all faults at the top of the result list such that the faults with the least lines come first. With the high number of faultless lines compared to faulty ones, the optimal AUCEC approaches zero.
As the optimal curve depends on the used granularity, the normalized result depends on it as well. Hence, these normalized AUCEC values can not directly be compared between studies using different granularities.

As mentioned, AUCEC values are often calculated for only parts of the $x$-axis, e.g., \texttt{AUCEC1} is the area up to $1$\% of LOC and \texttt{AUCEC5} up to $5$\% of LOC\@.
The reasoning behind focusing on early parts of result lists relies on results from Parnin and Orso~\cite{parninAreAutomatedDebugging2011a}, as well as Long and Rinard~\cite{longAnalysisSearchSpaces2016}. They found that developers would only inspect the first few elements of a result list, and even automated tools performed best with result numbers in the low hundreds.

\subsubsection{$E_{\mr{inspect}}$}~\cite{zouEmpiricalStudyFault2019}. $E_{\mr{inspect}}$ measures the expected rank of the first faulty element in a ranked list.
Assuming a group of $t$ tied elements starting at $P_{\mr{start}}$ that contains $t_f$ faulty elements and there is no faulty element before $P_{\mr{start}}$, $E_{\mr{inspect}}$ is defined as:

\begin{equation}
E_{\mr{inspect}} = P_{\mr{start}} + \sum_{k=1}^{t-t_f}k\frac{\binom{t-k-1}{t_f-1}}{\binom{t}{t_f}}    
\end{equation}

Based on $E_{\mr{inspect}}$ and the \texttt{acc@n} measure by Le et al.~\cite{b.leLearningtorankBasedFault2016a}, $E_{\mr{inspect}}@n$ counts the number of faults that were within the top $n$ positions of the ranked list.
A value of $10$ is reasonable for users, as most users will only inspect the first few entries~\cite{parninAreAutomatedDebugging2011a}.
For automated tools, Long and Rinard~\cite{longAnalysisSearchSpaces2016} propose $200$, which they found to work well.

We propose the $E_{\mr{inspect}}F$ evaluation metric as the lowest $E_{\mr{inspect}}$ value of all faults or, the absolute number of LOC that has to be inspected to encounter the first fault.
If a developer inspects the first $10$ elements of the result list, that is the threshold to meet. The same holds for automated tools.
While proportional evaluation metrics can be useful to compare different fault prediction metrics and models across multiple projects, we believe the absolute numbers are more relevant for most use cases.

\subsubsection{EXAM}~\cite{wongCrosstabbasedStatisticalMethod2008}.
\label{sec:exam}
The \texttt{EXAM} score is defined as the proportions of a project to be examined to find the first line of a fault (averaged across all faults).
Thus, an algorithm with a lower exam score finds faults faster, on average, than one with a higher exam score.
In terms of $E_{\mr{inspect}}$, the \texttt{EXAM} score can be defined as,

\begin{equation}
\frac{1}{n}\sum_{i=1}^{n} E_{\mr{inspect}}(F_n), 
\end{equation}

\noindent where $F_n$ are the faults.

From our experience, the \texttt{EXAM} score is a good approximation for the \texttt{AUCEC100} with $\mr{EXAM} = 1 - \mr{AUCEC100}$ having a mean error of $0.005$ in our dataset.
For this reason, we only report \texttt{AUCEC1} and \texttt{AUCEC5}, and use the \texttt{EXAM} score instead of \texttt{AUCEC100}.

\subsubsection{Research question mapping.}
While the \texttt{AUROC} and \texttt{EXAM} values are averaging the performance across the entire result list, in realistic scenarios the performance in the early parts of the results are more important. This is why we also use the \texttt{AUCEC5}, \texttt{AUCEC1}, $E_{\mr{inspect}}100$, $E_{\mr{inspect}}10$, and $E_{\mr{inspect}}F$ measures.
These metrics allow us to answer the first research question more nuanced and with more certainty.

For the second research question, we measure the runtime of just Bugspots and Linespots, excluding the remaining parts of the evaluation code.\footnote{The evaluation suite's code, including metric calculation, is available together with the reference implementation~\cite{LinespotsReferenceImplementation}.}

\subsubsection{Comparing granularities.}
\label{sec:granularities}
As Bugspots and Linespots report their results on different granularity levels, we have to be careful when comparing them. One way to compare them is to transform the file-based results of Bugspots into line-based results, as proposed by Zou et al.~\cite{zouEmpiricalStudyFault2019}. We do this by setting each line's score to the corresponding file's score. 
This results in lists of ranked lines for both Bugspots and Linespots and, thus, all metrics can be calculated.
Depending on the exact way of how metrics are calculated, this kind of transformation can impact the results for Bugspots. We argue that this does not put Bugspots at a disadvantage, as we found that past research had calculated results so that files could only be inspected as a whole.
Transforming the results to the line granularity metrics like EXAM will show better performance for Bugspots due to the way blocks with the same score are handled.
While we do not think this will influence our conclusions, Bugspots might perform better with this kind of transformation.

While there could be value of a comparison between Linespots and Bugspots by transforming Linespots' results to file-level results, we found that the performance of Linespots was sensitive to the specific mapping used and thus decided against a comparison using the file-level granularity.

\subsection{Dataset}
\label{sec:dataset}

There are two options for what kind of dataset to use in the field of fault prediction, either a released repository of artifacts and faults or a self-built dataset based on repository mining.
Examples of released sets are PROMISE~\cite{shirabadPROMISERepositorySoftware2005} or Defects4J~\cite{justDefects4JDatabaseExisting2014}. The benefit of using such repositories is the ease of use, reliability, and the comparability they offer between studies using the same repositories.
The drawback, however, is the limitation of the content. Usually, only a limited number of projects, revisions, and information is available. Be it the limitation to a single programming language, a specific granularity, or lack of history and process information. 
Moreover, while faults in those repositories usually are verified, that can lead to a false sense of security, with artifacts marked as non-faulty. This could lead to an inflation of wrongly classified false positives as algorithms find real faults that are not marked as such in the repository.

The repository mining approach comes with a different set of trade-offs. While the dataset's size can be almost arbitrarily big, it is usually limited to open-source projects. The quantity of data is usually higher than for a pre-built dataset, but the quality might suffer. For this study, we wanted to build a representative sample for, at least, open-source projects. While there have been alternatives growing in recent years, GitHub remains the biggest collection of open source projects, which is why we collected our sample of projects from there.\footnote{$270\mr{m}+$ repositories on Github vs. $19\mr{m}+$ on GitLab at the time of writing.}
Ultimately, we applied a combination of different sampling strategies, as described by Baltes and Ralph~\cite{baltesSamplingSoftwareEngineering2020}.

First, we applied cluster sampling by randomly choosing projects from the  top-$1000$ starred projects on GitHub. While limited to the top-starred projects, we assume that this improves our generalizability.

We then used purposive sampling by collecting a number of projects from related studies such as the works of Rahman et al.~\cite{rahmanBugCacheInspectionsHit2011}, Tóth et al.~\cite{tothPublicBugDatabase2016}, D’Ambros et al.~\cite{dambrosExtensiveComparisonBug2010} and Zou et al.~\cite{zouEmpiricalStudyFault2019}.
These projects would allow us to better connect this study's results with the broader field of fault prediction and localization.
To those, we also added additional projects to receive a more disparate sample, e.g., from the web-commerce domain.

All of the projects were filtered to meet our requirements concerning commit count and concerning the reliability of commit messages.
As the reliability of commit messages is a threat to our validity, we then searched for projects that followed strict CMCs (while also enforcing this!) In the end, $32$ projects as shown in Table~\ref{tab:projects}, were added to the sample.

\subsubsection{Identification of fix-inducing commits.} To identify fix-inducing commits, we performed pattern matching on commit messages. While D'Ambros et al.~\cite{dambrosExtensiveComparisonBug2010} showed that simple string matching could be unreliable, we believe that is at least partially due to poorly chosen regular expressions (regexes), since we see evidence from other studies catching unnecessary false positives this way~\cite{wangVersionHistorySimilar2014, youmImprovedBugLocalization2017}.
To improve reliability, we used separate regexes for each project, derived from studying past commits (also shown in Table~\ref{tab:projects}). The table also shows that while the three CMCs differ in detail, they all use the same format to denote fix-inducing commits. 

\begin{savenotes}
\begin{table}
\caption{Project properties}
\label{tab:projects}
\centering
\tabsize
\begin{tabular}{|l|l|l|l|l|l|}
\toprule
Project & Language & Regex & CMC & Samples & Source \\
\midrule
angular & TypeScript & \textasciicircum fix(\textbackslash ((.*)\textbackslash ))?: & angular\footnote{\url{https://github.com/angular/angular/blob/22b96b9/CONTRIBUTING.md\#-commit-message-guidelines}} & 3 & CMC \\
angular-cli & TypeScript & \textasciicircum fix(\textbackslash ((.*)\textbackslash ))?: & angular & 2 & CMC \\
angular.js & JavaScript & \textasciicircum fix(\textbackslash ((.*)\textbackslash ))?: & angular & 3 & CMC \\
components & TypeScript & \textasciicircum fix(\textbackslash ((.*)\textbackslash ))?: & angular & 3 & CMC \\
discourse & Ruby & \textasciicircum fix(\textbackslash ((.*)\textbackslash ))?: & discourse\footnote{\url{https://meta.discourse.org/t/github-checkin-prefix-convention/19392}} & 3 & Random \\
electron & C++ & \textasciicircum fix(\textbackslash ((.*)\textbackslash ))?: & conventional\footnote{\url{https://www.conventionalcommits.org}} & 1 & CMC \\
karma & JavaScript & \textasciicircum fix(\textbackslash ((.*)\textbackslash ))?: & angular & 1 & CMC \\
material & JavaScript & \textasciicircum fix(\textbackslash ((.*)\textbackslash ))?: & angular & 2 & CMC \\
monica & PHP & \textasciicircum fix(\textbackslash ((.*)\textbackslash ))?: & conventional & 1 & CMC \\
umbrella & TypeScript & \textasciicircum fix(\textbackslash ((.*)\textbackslash ))?: & conventional & 2 & CMC \\
tinacms & TypeScript & \textasciicircum fix(\textbackslash ((.*)\textbackslash ))?: & conventional & 1 & CMC \\
bootstrap & JavaScript & fix  & None & 3 & Author \\
broadleafcommerce & Java & fix\textbar fixes\textbar fixed & None & 3 & Author \\
ceylon-ide-eclipse & Java & fix\textbar fixed  & None & 3 & Past Work \\
closure-compiler & Java & fix\textbar fixed  & None & 3 & Past Work \\
coala\footnote{The main author is a maintainer of coala} & Python & Fixes & None & 2 & Author \\
commons-math & Java & fix\textbar fixed  & None & 3 & Past Work \\
cpython & Python & fix\textbar fixes  & None & 3 & Author \\
evolution & C & fix\textbar fixes  & None & 3 & Past Work \\
ffmpeg & C & fix\textbar fixed  & None & 3 & Random \\
httpd & C & fix  & None & 3 & Random \footnote{httpd was chosen randomly, but is used in past work as well} \\
jfreechart & Java & fix\textbar fixed  & None & 1 & Past Work \\
junit5 & Java & fixes  & None & 3 & Past Work \\
lucene-solr & Java & fix  & None & 3 & Past Work \\
mongoose & JavaScript & fix:\textbar fix & None & 3 & Random \\
mysql-server & C++ & fix  & None & 3 & Author\\
prestashop & PHP & fix\textbar fixes  & None & 3 & Author \\
rails & Ruby & fix\textbar fixes  & None & 3 & Author \\
rt.equinox.framework & Java & fix  & None & 2 & Past Work \\
scikit-learn & Python & fix\textbar fixes  & None & 3 & Author \\
server\footnote{MariaDB Server} & C++ & fix\textbar fixed\textbar fixing  & None & 3 & Author \\
woocommerce & PHP & fix\textbar fixes  & None & 3 & Author \\
\bottomrule
\end{tabular}
\end{table}
\end{savenotes}

\subsubsection{Parameters.} We use a depth of $500$ commits for both algorithms as it is the default of the Bugspots reference implementation used by other studies~\cite{grigorikImplementationSimpleBug2019}. A preliminary investigation resulted in no improved performance for increased depths.
While Lewis et al.~\cite{lewisDoesBugPrediction2013} suggest tuning the weighting function for each project, we kept the default values to allow for better comparability with other studies using them. However, we calculate the ages of commits based on the index instead of the Unix timestamps. The use of timestamps resulted in some corner case problems with rewritten or merged histories. (We did not see a significant difference in results due to the change in time stamps.)
As our origins, the commit to run the algorithms from, we randomly chose up to three commits from the entire project history, under the condition that both the $500$ depth commits and the pseudo-future commits used for validation did not overlap between samples (i.e., stratified random sub-sampling).

\subsection{Validation data}
As we do not have a pre-built repository of faults to validate our predictions, we had to develop our own.
We used the commits after the origin commit as a pseudo future. We then used the same pattern matching and line tracking as Linespots uses, to identify lines that already existed at the origin commit and were later removed during a fix-inducing commit.
This follows Pearson et al.~\cite{pearsonEvaluatingImprovingFault2016} and Le et al.~\cite{b.leLearningtorankBasedFault2016a}, in that faulty lines are those that get removed; remember git does not modify lines during a fix-inducing commit.
In case only new lines are added, Pearson et al.~\cite{pearsonEvaluatingImprovingFault2016} proposes to tag the line immediately following the newly added lines as faulty.
Based on this collection of faulty lines, we could calculate the evaluation metrics.

\subsubsection{Determining when a fault was predicted.} The last missing piece for evaluation is to decide at what point a fault counts as predicted. With rougher granularities like modules, it is rather straightforward. If you propose a module it can either be faulty or not, although faults involving multiple modules would pose problems.

It becomes even more complicated with lines, as most faults will consist of multiple lines, and it is not clear if evaluating on line- or fault-level is the right choice if such a thing even exists.
Following Zou et al.~\cite{zouEmpiricalStudyFault2019}, we evaluate on the fault level, determining how well the algorithms predicted faults instead of individual faulty lines.
Both Pearson et al.~\cite{pearsonEvaluatingImprovingFault2016} and Meng et al.~\cite{mengSystematicEditingGenerating2011} argue that a fault is predicted if any individual line is predicted.
An alternative was proposed by Rahman et al.~\cite{rahmanComparingStaticBug2014} in the form of partial credit, where for each line, a partial prediction credit is given. This might be a more honest representation of performance.
Similarly, Pearson et al.~\cite{pearsonEvaluatingImprovingFault2016} present their results for the best case, average case, and worst-case scenario. The best case is similar to what we use, where a fault counts as predicted with just a single line. The worst-case requires all lines for a fault to count as predicted while the average requires half of the lines.

While the idea of giving partial credit, or presenting performance with different thresholds, could paint a more balanced picture of performance, it is not compatible with all fault prediction metrics that are commonly used. Furthermore, this can add substantial computational requirements, especially when working on line-level granularity.

\subsection{Analysis}
\label{sec:analysis}
To analyze Bugspots' and Linespots' effects on the evaluation metrics, we use Bayesian data analysis as outlined by Schad et al.~\cite{schadPrincipledBayesianWorkflow2020}.
By doing so, we follow the recommendation to move away from $p$-value based reporting as called for by Wasserstein et al.~\cite{wassersteinMovingWorld052019}.
In this paper, we only give a summary of the process. Documenting the entire process in detail would exceed this paper's scope; thus, we instead provide a replication package and detailed analysis reports~\cite{scholzLinespotsAnalysisRepository2019}.

\begin{figure}
\centering
   \includegraphics[width=0.6\textwidth]{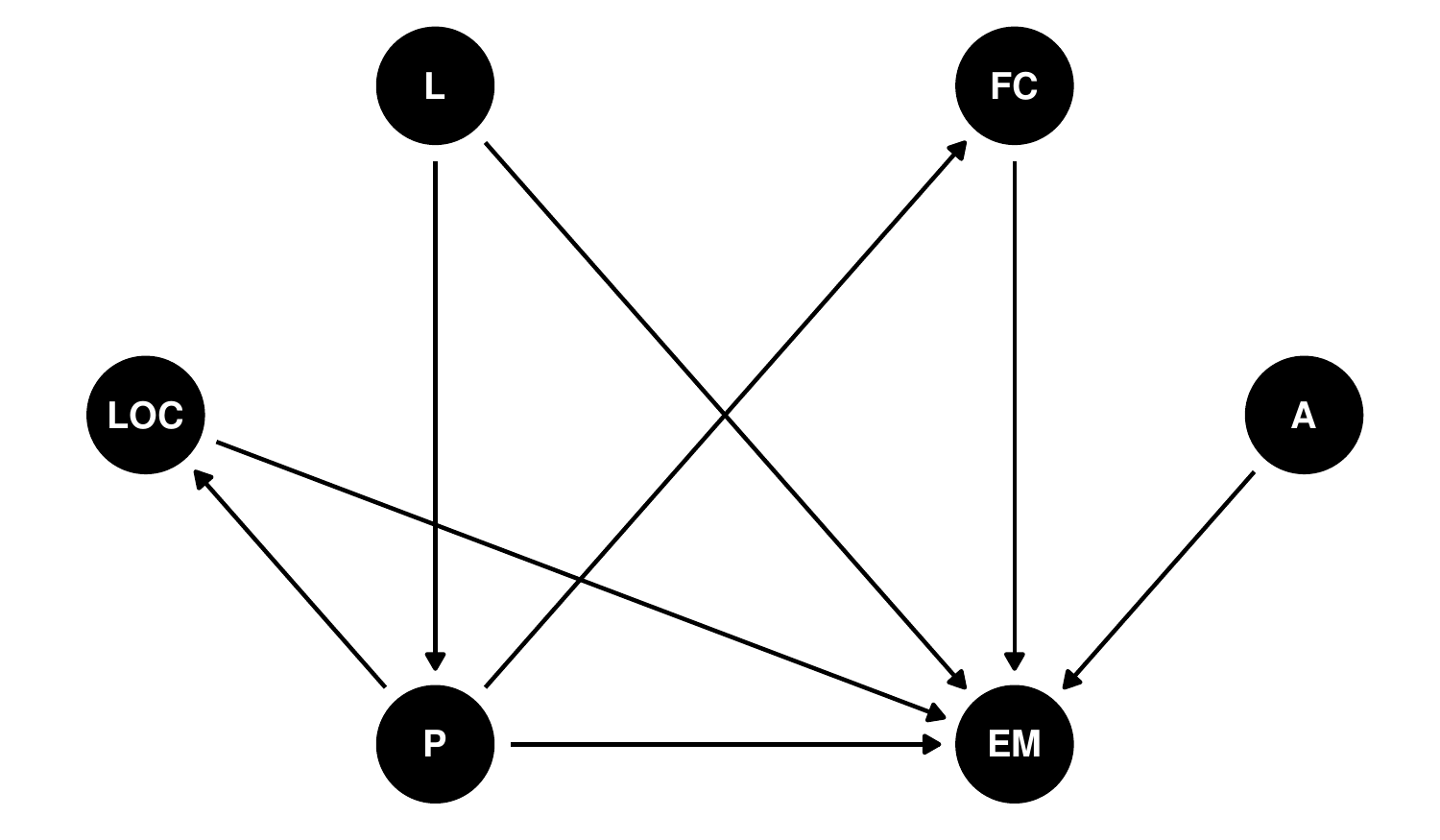}
    \caption{Simplified DAG of the experiment. A: algorithm, EM: evaluation metrics, FC: fix count, L: programming language, LOC: lines of code, P: project}
    \label{fig:dag}
\end{figure}

Based on the work of Pearl~\cite{pearlSevenToolsCausal2019, pearlCausalityModelsReasoning2009}, we started by building a Directed Acyclic Graph (DAG) to represent the causal relationships of our measured variables, as shown in Figure~\ref{fig:dag}.
This not only allows us to make more stronger claims about the causality but also documents our assumptions regarding the process. This allows readers to better understand our reasoning, find potential flaws in it and adapt it for their own work.
The arrows in the graph display a causal relationship between the nodes and can be used to check for confounders and guide model building.
We can then query the graph and use \textit{do}-calculus to check if the query can be answered.
Below is the query for a statistical model that tries to estimate the algorithms' causal effect on the evaluation metrics controlling for project, language, LOC, and fix count. Or more precise, the probability of EM given treatment A and controlling for P, L, LOC and, FC\@.

$$P(\mr{EM}|\mr{do}(\mr{A}), \mr{P}, \mr{L}, \mr{LOC}, \mr{FC})$$

The \textit{do}-operator indicates an intervention or treatment; in our case, the use of either Bugspots and Linespots. The goal of the \textit{do}-calculus application is to eliminate the \textit{do}-operator, so we are left with classical probability expressions.
In our case, we will apply the second rule of \textit{do}-calculus that states, 

$$P(\mr{Y}|do(\mr{X}), \mr{Z}) = P(\mr{Y}|\mr{X}, \mr{Z})$$ 

\noindent if Z satisfies the back-door criterion. The back-door criterion is satisfied if a set Z of variables blocks all back-door paths from X to Y.
Back-door paths are paths that start with an arrow pointing into X (algorithm in our case).
With only one outgoing arrow for algorithm, there are no possible back-door paths in the graph. Hence, the backdoor-criterion is satisfied for any combination of the four control variables. If we apply the rule to our query, we get the following,

$$P(\mr{EM}|\mr{do}(\mr{A}), \mr{P}, \mr{L}, \mr{LOC}, \mr{FC}) = P(\mr{EM}|\mr{A}, \mr{P}, \mr{L}, \mr{LOC}, \mr{FC})$$

With no \textit{do}-operator left, we confirm that we can measure the causal effect of the algorithm on the evaluation metrics controlling for project, language, LOC, and fix count. (Disregarding omission bias, which we will discuss later.) 

This technique contrasts with the practice of just adding all available predictors to a model, and reduces the risk of confounding and bias, based on the assumptions. For a good primer on the use of DAGs, we recommend~\cite{mcelreathStatisticalRethinkingBayesian2020}.

With the DAG's possible models, we then chose the likelihood according to the assumed underlying data generation process. Finally, we sampled from the models using Hamiltonian Monte Carlo. Concerning priors, which we set on the model, we aimed to follow recommended practices. We verified that the combinations of priors were uniform on the outcome scale by conducting prior predictive checks.

After compiling and sampling multiple models per evaluation metric using \texttt{brms}~\cite{burknerBrmsPackageBayesian2017a} and Stan~\cite{carpenterStanProbabilisticProgramming2017}, we ensured all necessary model diagnostics passed and then compared models' relative out of sample prediction capabilities (for details please see the replication package)~\cite{vehtariLooEfficientLeaveOneOut2019}. The results we report are from the models that showed the best relative out of sample prediction capabilities.

Below is the definition of one of the models used in our analysis. We will next use it to explain our approach.

\setcounter{equation}{0}
\begin{IEEEeqnarray}{rClr}
    \mr{EXAM_i} & \thicksim & \mr{Beta}(\mu_i, \phi) & \\
    \logit(\mu_i) & = & \alpha + \beta_{A} \mr{Algorithm}_i + \beta_{L} \mathrm{LOC}_i & \\
    & + & \beta_F \mr{FixCount}_i + \alpha_{\mr{PROJECT}[i]} & \\
    \alpha & \thicksim & \mr{Normal(-1, 1)} & \\
    \beta_A, \beta_L, \beta_F & \thicksim & \mr{Normal}(0, 0.15) & \\
    \alpha_{\mr{PROJECT}} & \thicksim & \mr{Normal(0, \phi_p)} &\textrm{for PROJECT} = 1,\ldots,32 \\
    \phi_p & \thicksim & \mr{Weibull}(2,1) &\\
    \log(\phi) & \thicksim &\ \mr{Normal(50, 20)} &
\end{IEEEeqnarray}

On Line $1$, we define the use of a Beta distribution to model the individual exam scores. We use a Beta distribution, as exam scores can be a real number $(0,1)$. We further use a parameterization of the beta distribution with a mean $\mu_i$ and a precision parameter $\phi$~\cite{burknerBrmsPackageBayesian2017a}.
The next two lines are the linear model that we use to approximate $\mu_i$. As the Beta distribution's mean has to be $(0,1)$, we use a $\logit$ link function to transform the results of the linear model to the $(0,1)$ interval.

In the linear model, we then use a global intercept $\alpha$, slopes for algorithm ($\beta_A$), LOC ($\beta_L$), and the fix count ($\beta_F$). Finally, we have varying intercepts per project. The idea is that we will model a deviation from the global intercept for each project. Some projects we will be able to estimate quite precisely and other projects will then learn something from these projects (a concept called partial pooling, used mainly to avoid overfitting, i.e., learn too much from the data).

Line $4$ is the prior for the global intercept, in this case, a normal distribution with mean $-1$ and standard deviation of $1$. We derived the $-1$ mean from experience with our past work, where the mean EXAM value was around $0.226$ (i.e., approximately $-1$ on the logit scale).

The priors on Lines $5$, $6$ and $7$ were chosen iteratively using prior predictive checks to allow the model to explore the outcome space uniformly. Lines $6$ and $7$ define the hierarchical or multi-level part of the model, where we define varying intercept per project and assume all project intercepts to have a mean of 0 and a Weibull$(2,1)$ distributed standard deviation.
The prior for $\phi$ on Line $7$ uses a $\log$ link by default, and we set a wide prior, as we assume most of the outcome values to be small and somewhat concentrated, which leads to higher $\phi$ values on the $\log$ scale.

 In this section we explained our choice of evaluation metrics, the approach for how we built our dataset and validation data and gave a brief introduction to the usage of DAGs and Bayesian modeling. We will explain how to interpret the results of this kind of analysis in the following results section.

\section{Results}
\label{sec:results}

Before presenting the results, we want to provide some details on how to interpret the results, as the use of Bayesian data analysis is a novelty in the field of fault prediction, as far as we know.
We present interesting effects in two ways: 1. the effects of a model as-is and, 2. as conditional effects.

While, usually, one can just present model effects as-is, some link functions increase the possibility for misinterpretations.
For all models where the outcomes range between $0$ and $1$, we use Beta likelihoods and a $\logit$ link function, to map the result of the linear model to the $p \in (0,1)$ scale. However, the effect a parameter has on the outcome scale depends on where the model lies on the $\logit$ scale (as it is not linear).

To illustrate this, assume $\mu=0$ on the $\logit$ scale, which is equal to $0.5$ on the outcome scale.
With an effect of $-1$, the result becomes $-1$ on the $\logit$ or $0.27$ on the outcome scale---a difference of $0.23$.
However, if $\mu=-2$ on the $\logit$ scale, or $0.12$ on the outcome scale, and we have the same effect of $-1$, the result equals $-3$ on the $\logit$ scale or $0.05$ on the outcome scale. Now the effect only leads to a difference of $0.07$ on the outcome scale.

\begin{figure}
    \centering
    \includegraphics[width=0.7\textwidth]{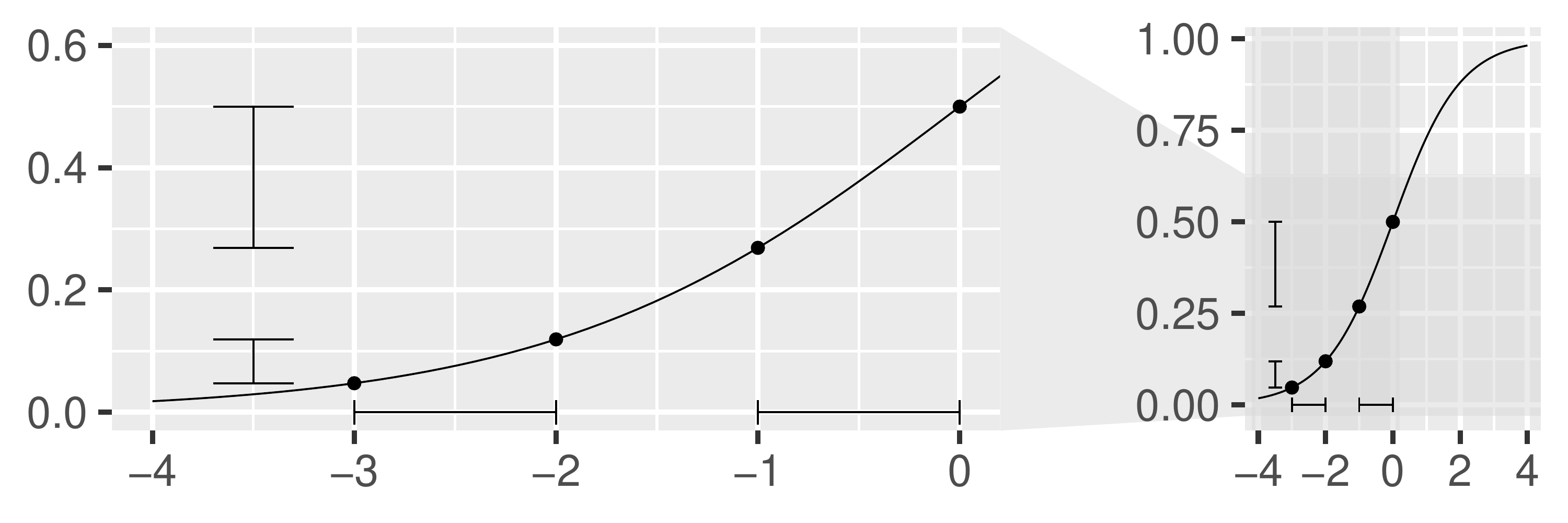}
    \caption{Non-linear transformation of the logistic function}
    \label{fig:logistic}
\end{figure}

Figure \ref{fig:logistic} shows a logistic curve with the example points. It shows how the $y$-axis's differences become smaller with a larger distance from $0$ on the x-axis.
So while effects on the logit scale can still be used to make statements such as ``Algorithm $A$ has a positive effect on metric $M$'', the size of that effect on the outcome scale is not as intuitive as when designing models employing an identity link.
(Some of the models use a $\log$ link instead of $\logit$, and while the $\log$ link does not behave exactly like the $\logit$, it shares the nonlinearity.) 

One pragmatic approach to solve this are conditional effects. Conditional effects set all model predictors, besides the one we are interested in, to their mean value or reference category for factors. This allows us to see the effect of the predictor of interest on the outcome scale for the average sample, i.e., in our case, this tells us the effect of the algorithm on the average project. Worth mentioning in this context is that the effect of interest very often has broad uncertainty connected to its interpretation. This is due to the uncertainty propagated by the model, and, generally speaking, we consider it to be a good thing.
The conditional effects plots we show in this paper show the median and 95\% credible intervals.
A summary of the conditional effects for all evaluation metrics is shown in Table~\ref{tab:conditional-effects-summary}.

The replication package~\cite{scholzLinespotsReproducabilityDocker2020} and analysis report~\cite{scholzLinespotsAnalysisRepository2019} allow the reader to reproduce both the data collection and analysis procedure.

\subsection{RQ1: Predictive performance comparison}
\begin{figure}
    \centering
    \subfloat[Logit scale effect of Bugspots]{
        \includegraphics[width=0.45\textwidth]{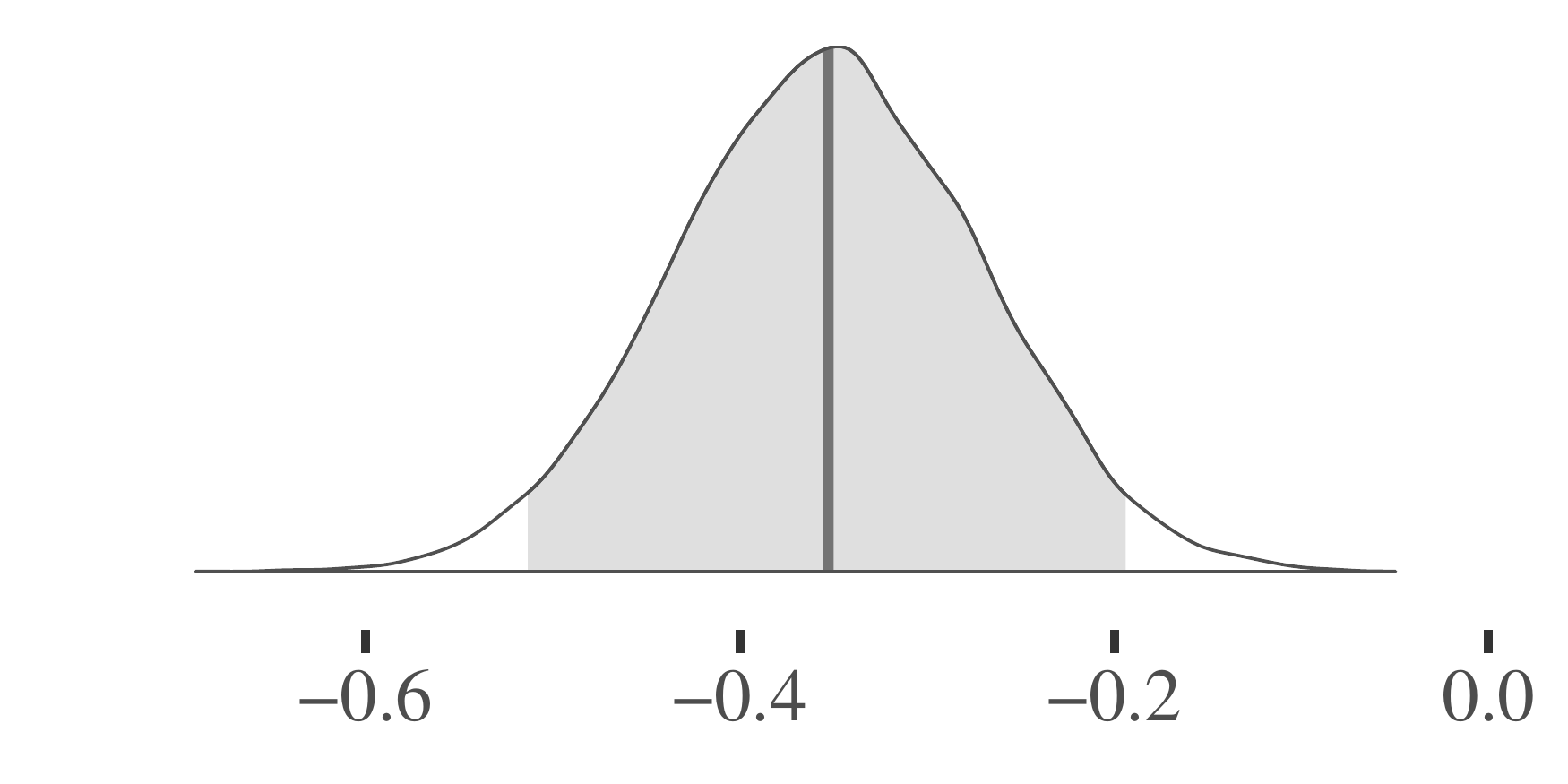}
    }
    \hfill
    \subfloat[Conditional effects]{
        \includegraphics[width=0.45\textwidth]{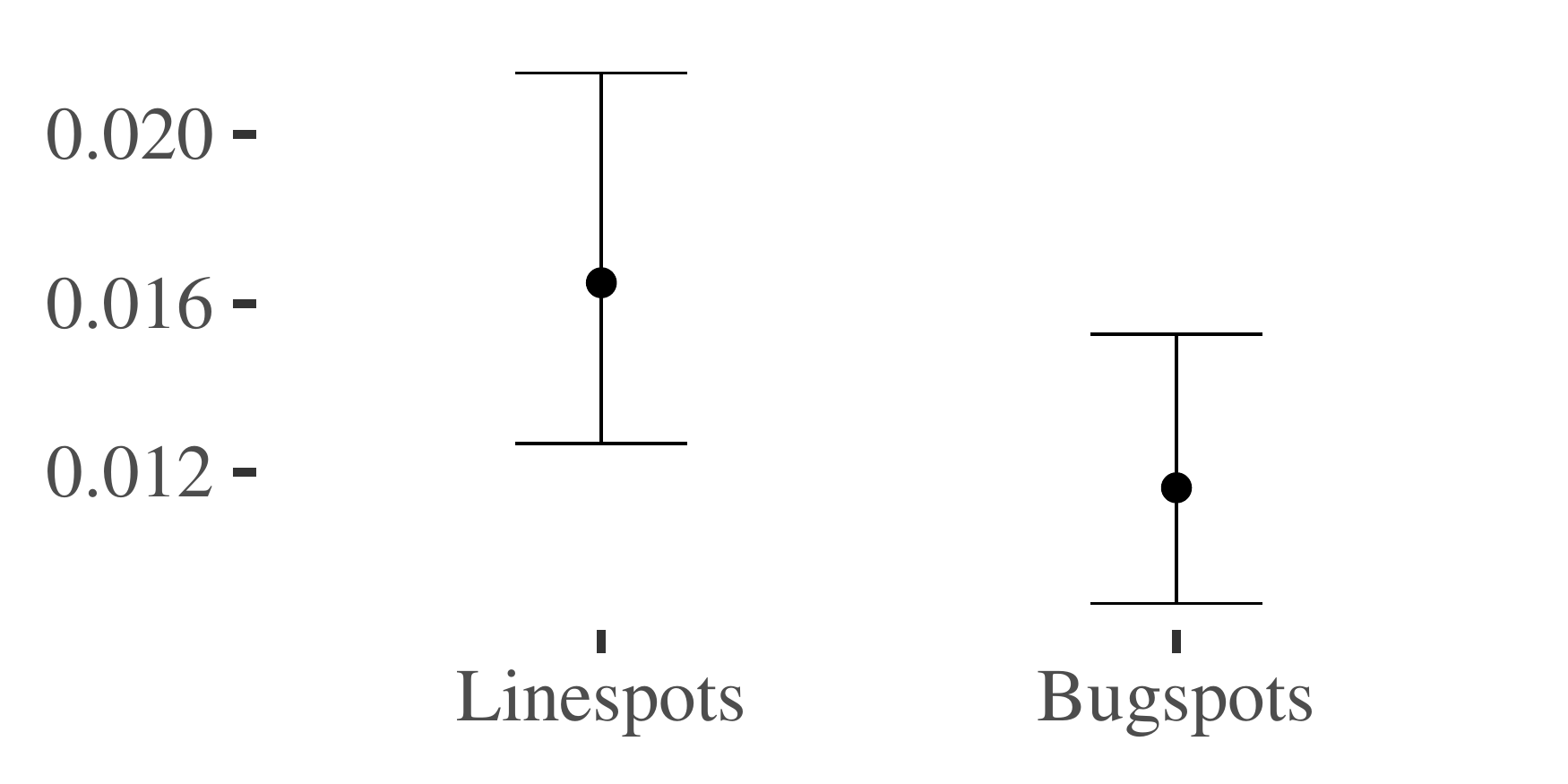}
    }
    \caption{Effects of algorithm on \texttt{AUROC}}
    \label{fig:res-auroc}
\end{figure}

To answer our research question, we collected seven evaluation metrics.
We start with \texttt{AUROC} and \texttt{EXAM} as more high level, averaging evaluation metrics.
Figure~\ref{fig:res-auroc} shows the effect of Bugspots compared to Linespots on the $\logit$ scale and the conditional effects of both algorithms with median and 95\% credible intervals. While there is overlap between the two algorithms on the outcome scale (b), the effect on the logit scale (a), shows us that Bugspots produces lower \texttt{AUROC} values than Linespots. The uncertainty in the conditional effects is caused by the uncertainty propagated by the model, and not the necessarily by the algorithm.

\begin{figure}
    \centering
    \subfloat[Logit scale effect of Bugspots]{
        \includegraphics[width=0.45\textwidth]{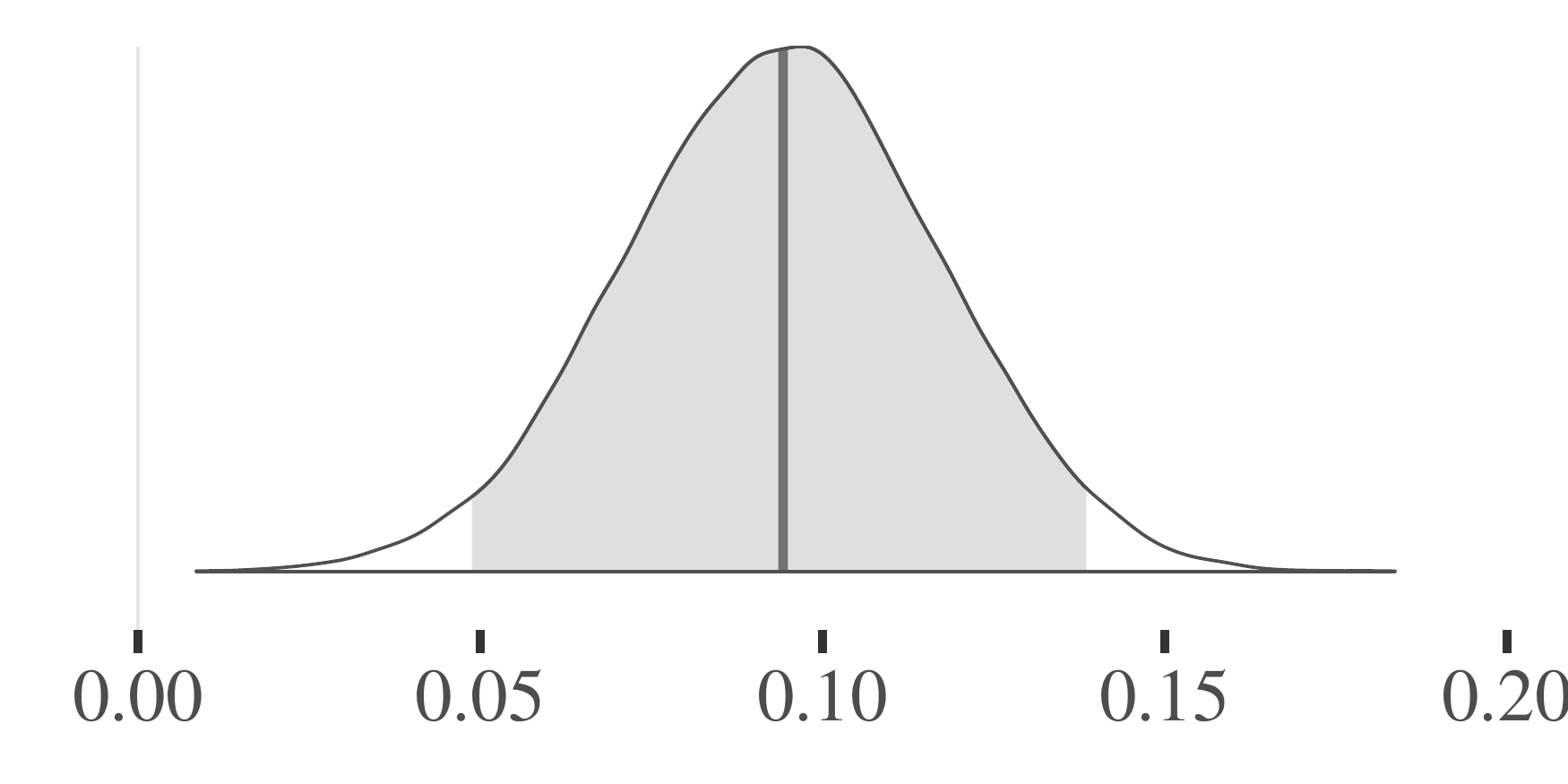}
    }
    \hfill
    \subfloat[Conditional effects]{
        \includegraphics[width=0.45\textwidth]{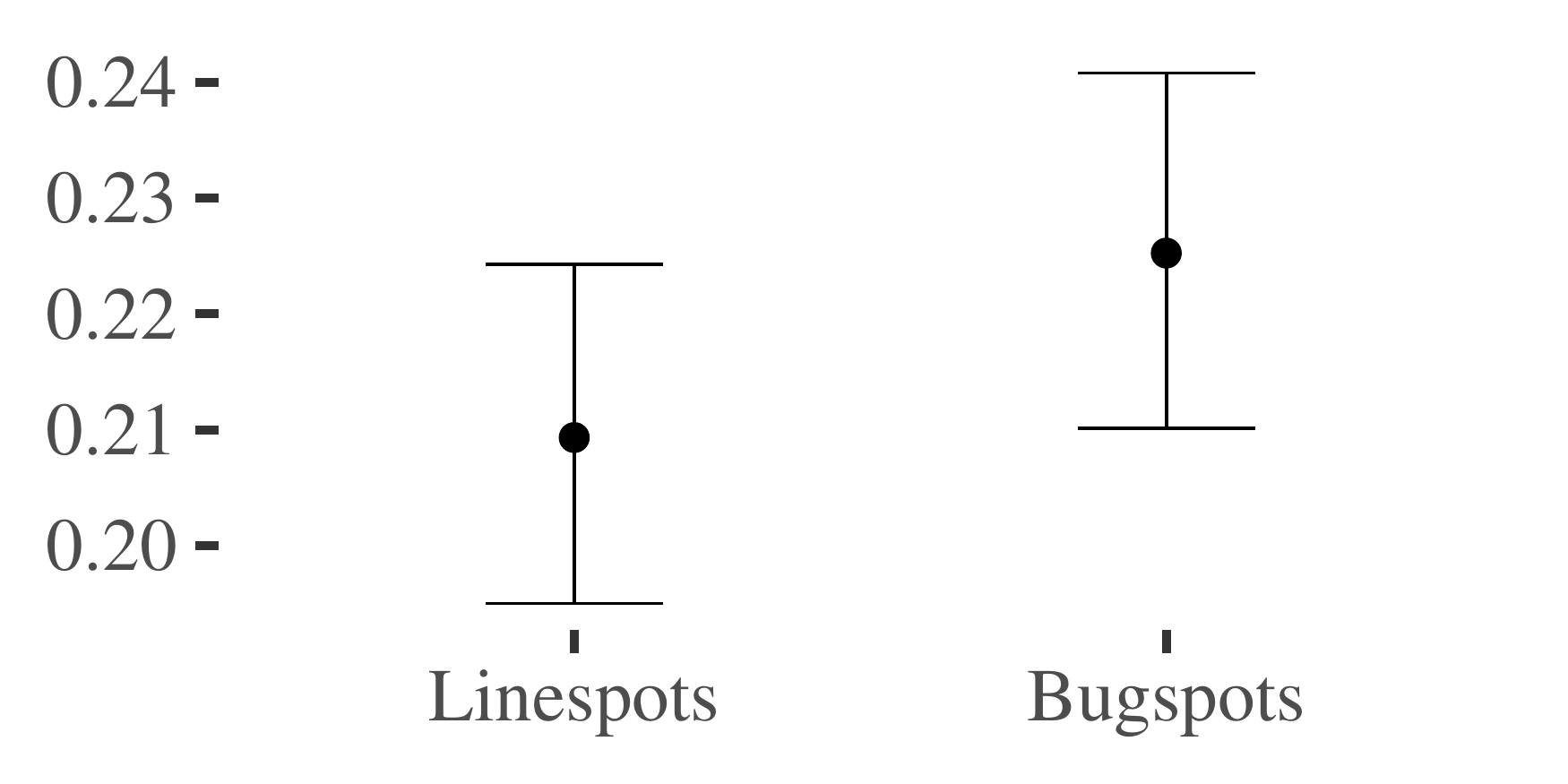}
    }
    \caption{Effects of algorithm on \texttt{EXAM}}
    \label{fig:res-exam}
\end{figure}

The \texttt{EXAM} score shows a similar picture in Figure~\ref{fig:res-exam}. Again, the difference between Linespots and Bugspots on the $\logit$ scale is entirely positive, while there is overlap on the outcome scale. As in the previous example, this points towards Linespots producing lower \texttt{EXAM} scores than Bugspots.

\begin{figure}
    \centering
    \subfloat[Logit scale effect of Bugspots]{
        \includegraphics[width=0.45\textwidth]{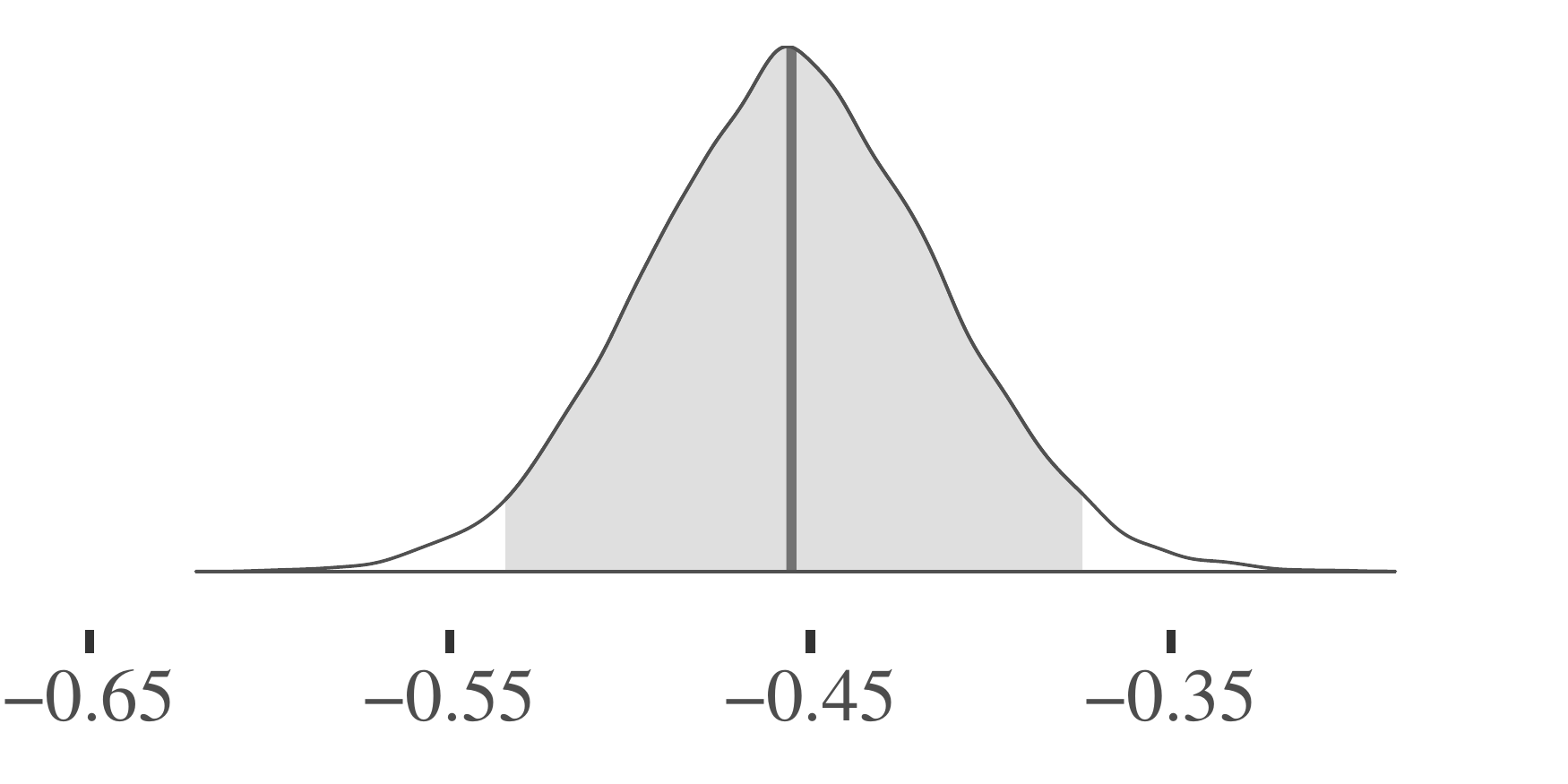}
    }
    \hfill
    \subfloat[Conditional effects]{
        \includegraphics[width=0.45\textwidth]{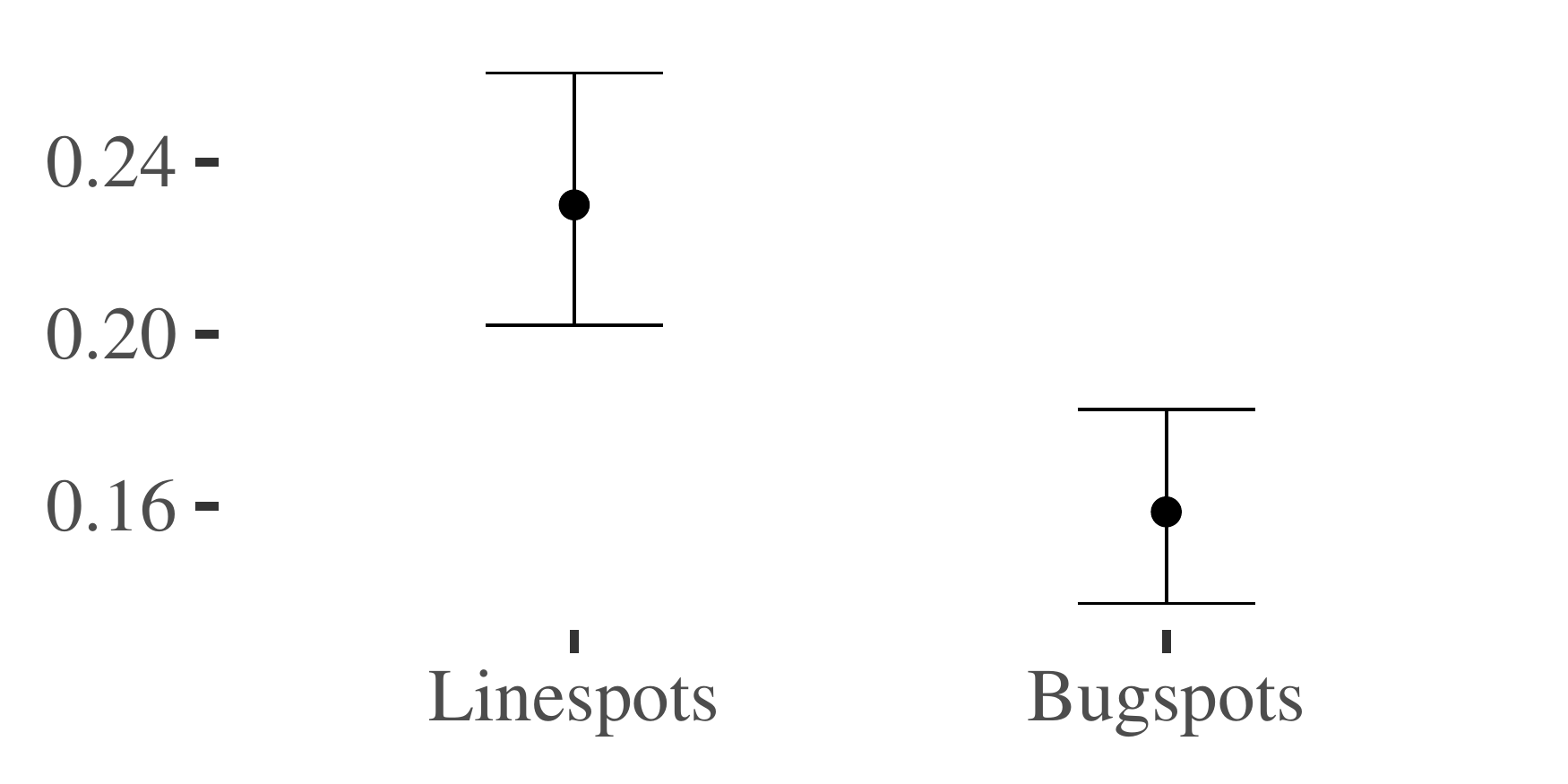}
    }
    \caption{Effects of algorithm on \texttt{AUCEC5}}
    \label{fig:res-aucec5}
\end{figure}

\begin{figure}
    \centering
    \subfloat[Logit scale effect of Bugspots]{
        \includegraphics[width=0.45\textwidth]{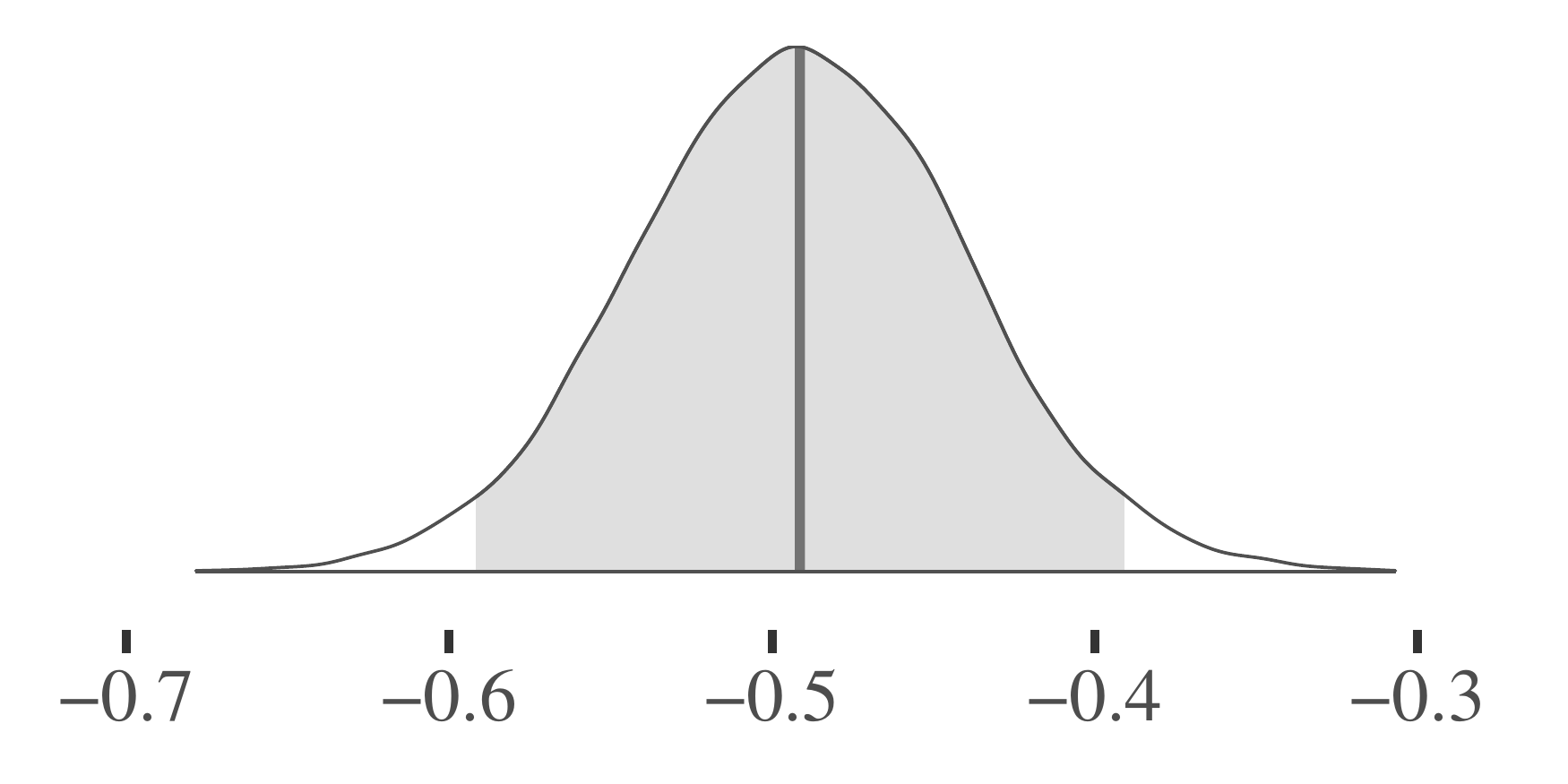}
    }
    \hfill
    \subfloat[Conditional effects]{
        \includegraphics[width=0.45\textwidth]{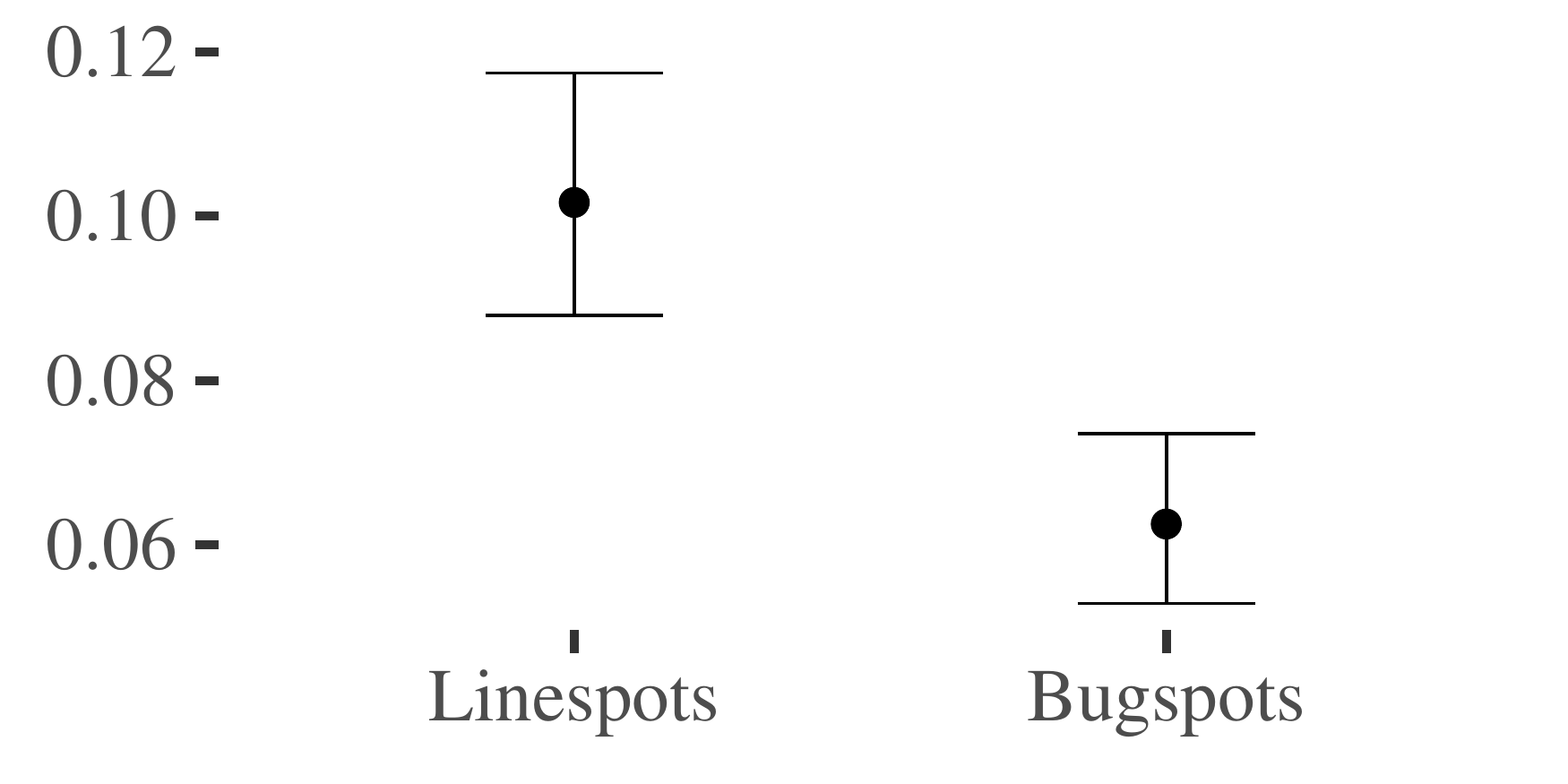}
    }
    \caption{Effects of algorithm on \texttt{AUCEC1}}
    \label{fig:res-aucec1}
\end{figure}

The \texttt{AUCEC5} results in Figure~\ref{fig:res-aucec5}, and \texttt{AUCEC1} results in Figure~\ref{fig:res-aucec1}, are more focused towards the early parts of the result lists.

\texttt{AUCEC1} and \texttt{AUCEC5} show apparent negative effects for Bugspots on the $\logit$ scale, with no overlap for the algorithms on the outcome scale. This is a clear indicator that Linespots consistently produces higher \texttt{AUCEC1} and \texttt{AUCEC5} values than Bugspots.

\begin{figure}
    \centering
    \subfloat[Log scale effect of Bugspots]{
        \includegraphics[width=0.45\textwidth]{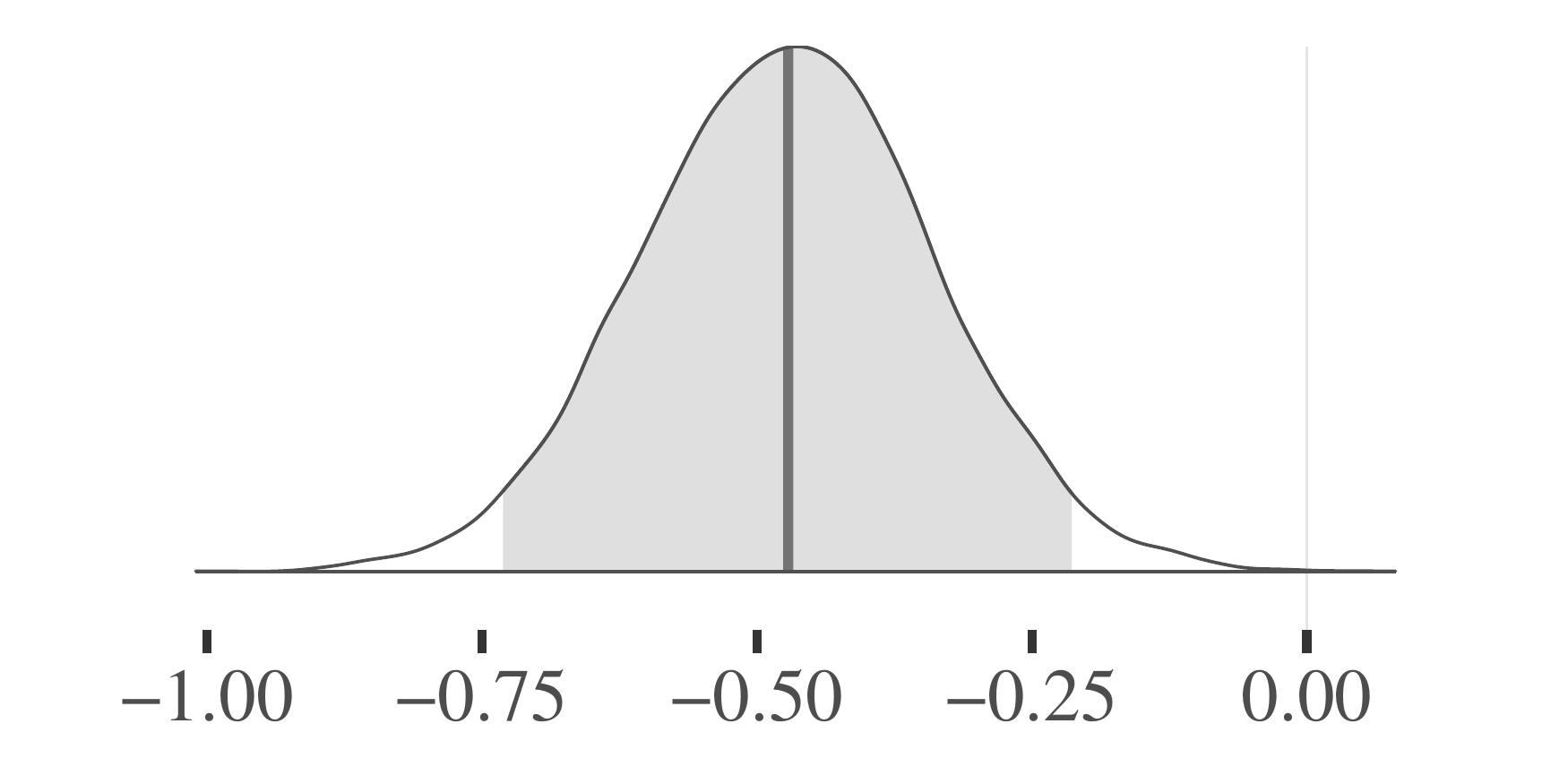}
    }
    \hfill
    \subfloat[Conditional effects]{
        \includegraphics[width=0.45\textwidth]{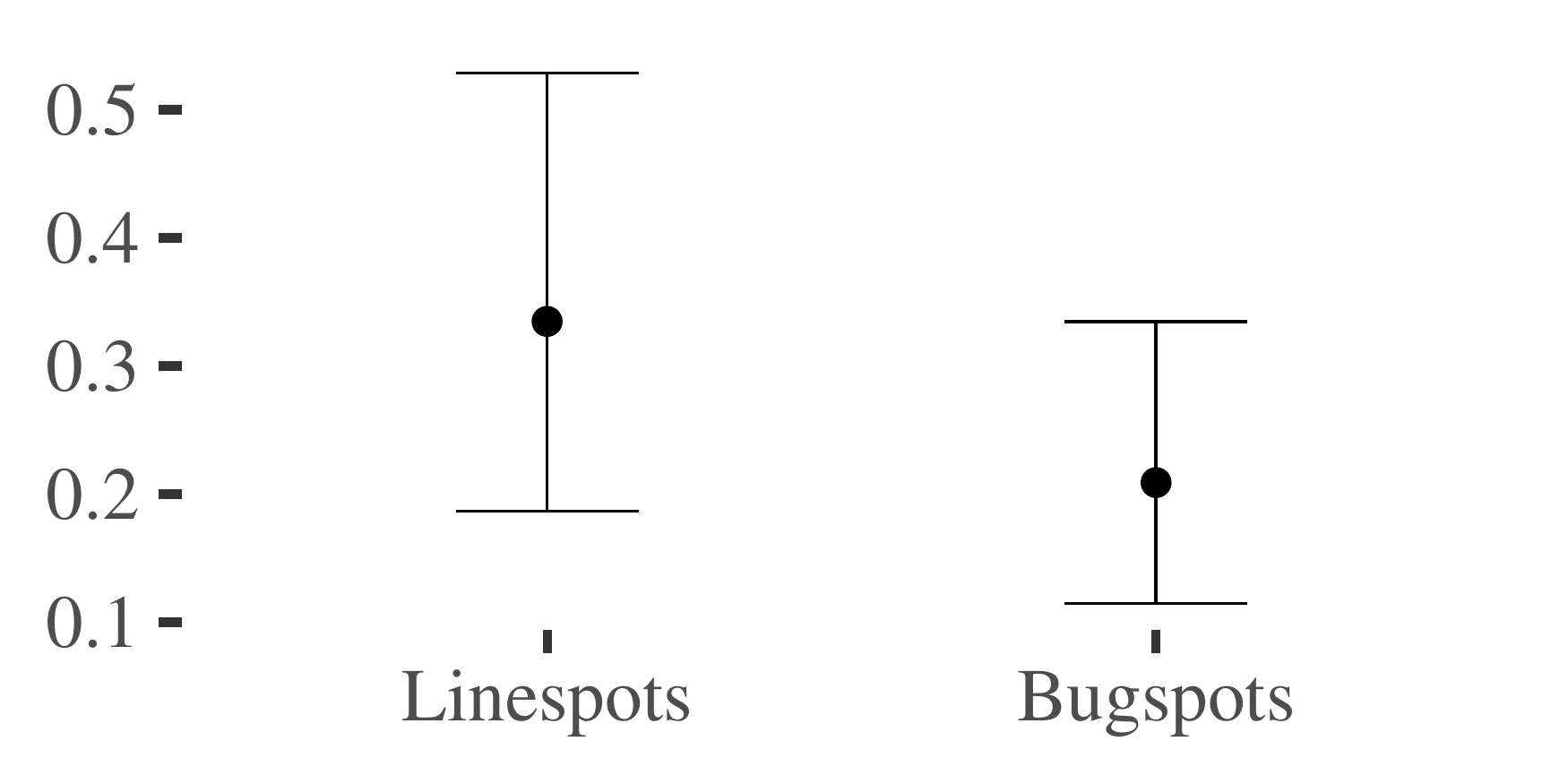}
    }
    \caption{Effects of algorithm on $E_{\mr{inspect}}@100$}
    \label{fig:res-einspect100}
\end{figure}

\begin{figure}
    \centering
    \subfloat[Log scale effect of Bugspots]{
        \includegraphics[width=0.45\textwidth]{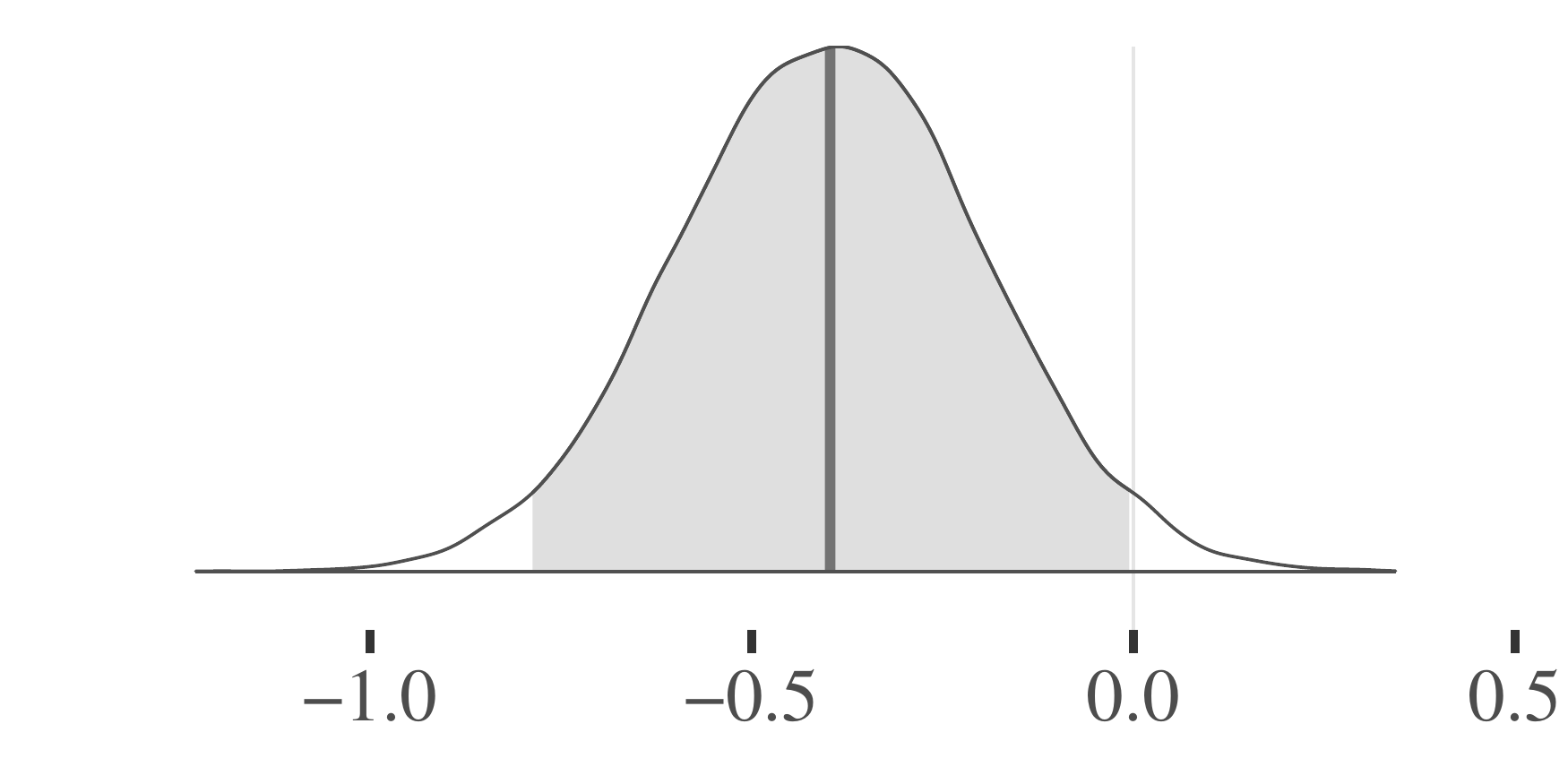}
    }
    \hfill
    \subfloat[Conditional effects]{
        \includegraphics[width=0.45\textwidth]{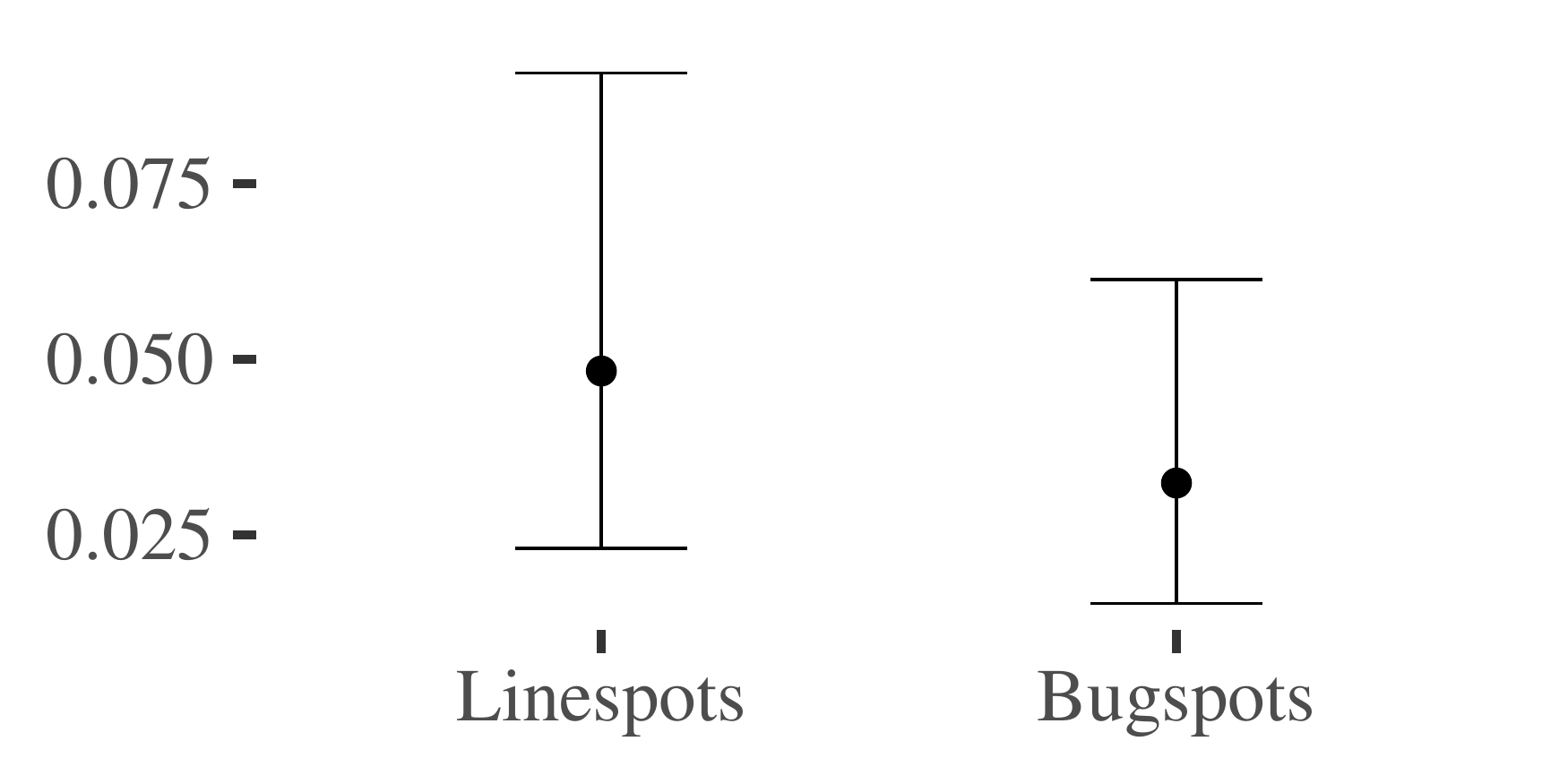}
    }
    \caption{Effects of algorithm on $E_{\mr{inspect}}@10$}
    \label{fig:res-einspect10}
\end{figure}

Next, we look at the very first parts of the result lists with the $E_{\mr{inspect}}@10$ and $E_{\mr{inspect}}@100$ evaluation metrics.
The results of $E_{\mr{inspect}}@10$, see Figure~\ref{fig:res-einspect10}, show some overlap with $0$ for the $\logit$ effect of Bugspots, while the outcome scale shows much overlap. This is probably caused by the zero-inflation that both algorithms produce, i.e., their overall low performance.
However, Linespots can produce higher $E_{\mr{inspect}}@10$ values than Bugspots and does so on average.
The results for $E_{\mr{inspect}}@100$, see Figure~\ref{fig:res-einspect100}, are similar, albeit more distinct. On the $\logit$ scale, there is just barely some overlap in the tail, and the overlap on the outcome scale is less pronounced.
There is still some uncertainty here, but the trend seems to be for Linespots to outperform Bugspots in $E_{\mr{inspect}}@100$.

\begin{figure}
    \centering
    \subfloat[Log scale effect of Bugspots]{
        \includegraphics[width=0.45\textwidth]{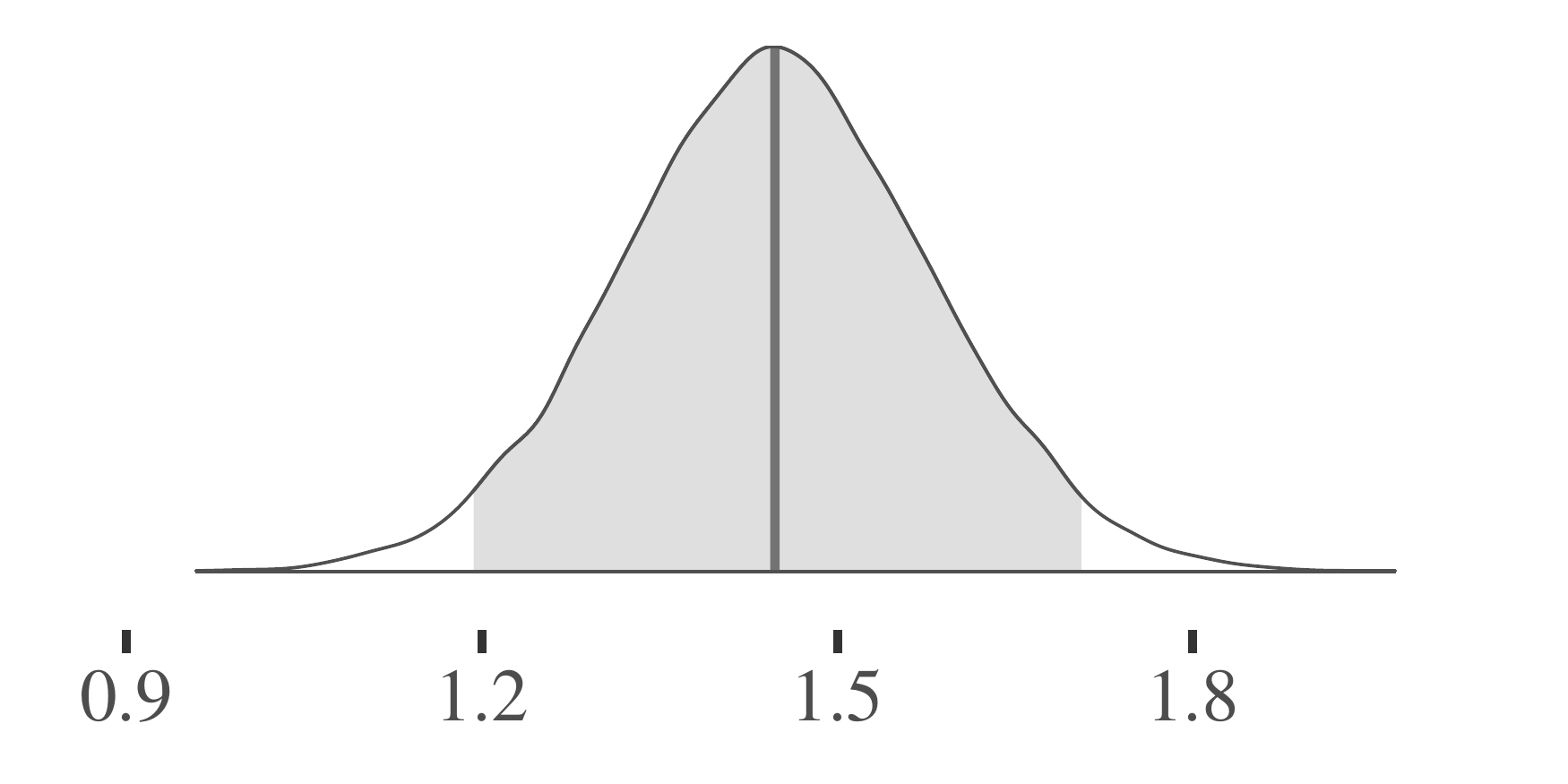}
    }
    \hfill
    \subfloat[Conditional effects]{
        \includegraphics[width=0.45\textwidth]{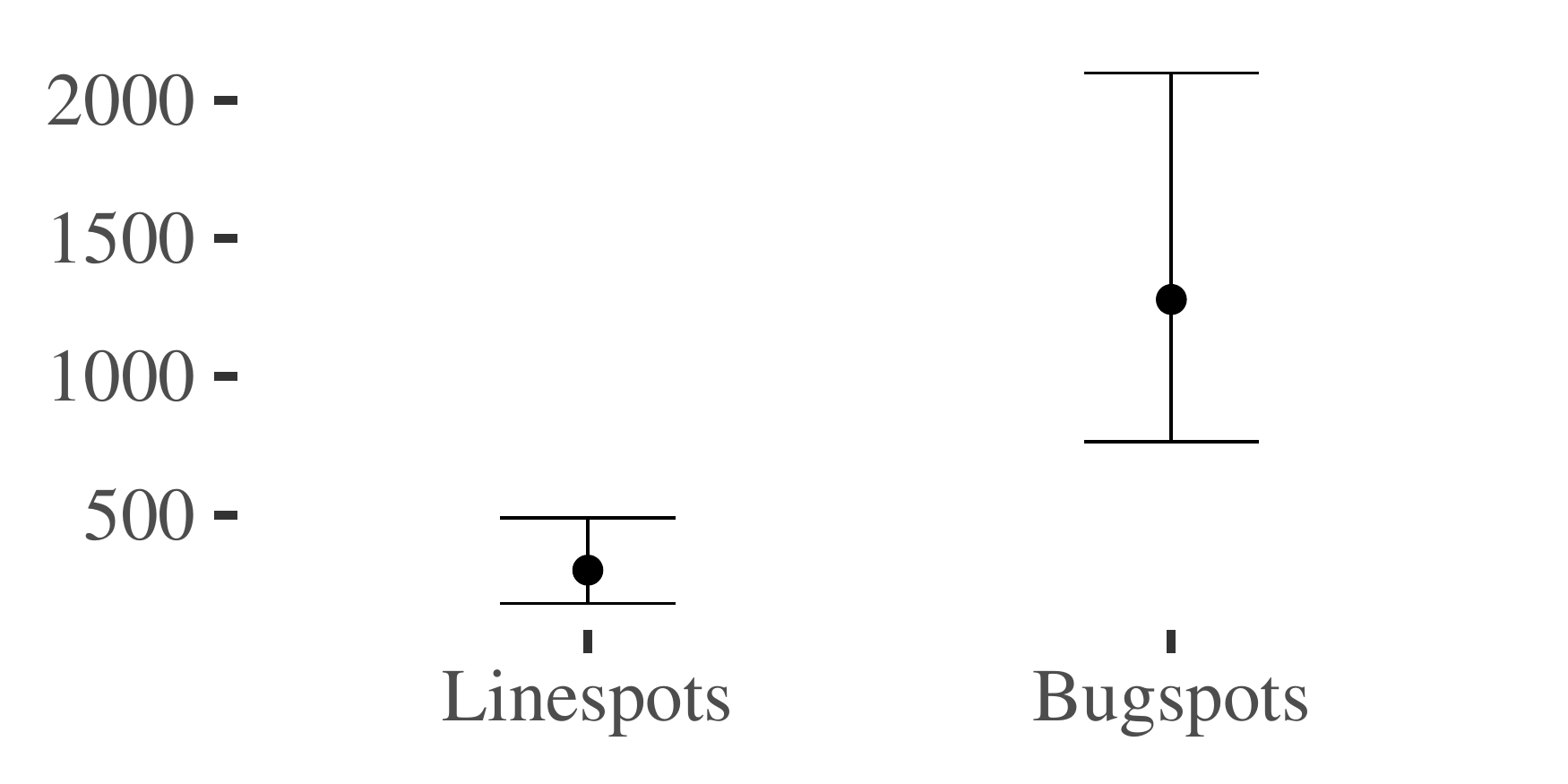}
    }
    \caption{Effects of algorithm on $E_{\mr{inspect}}F$}
    \label{fig:res-einspectf}
\end{figure}

Finally, we compared the position of the first fault in the result list with the $E_{\mr{inspect}}F$ scores.
Again, the results in Figure~\ref{fig:res-einspectf}, indicate that the effect of Bugspots on the $\logit$ scale is entirely positive, and there is no overlap on the outcome scale. Linespots consistently produces lower $E_{\mr{inspect}}F$ scores than Bugspots.

\begin{table}
\caption{Conditional effect size summary}
\label{tab:conditional-effects-summary}
\centering
\tabsize
\begin{tabular}{ c c c c c c }
\toprule
Algorithm & Mean & Median & SE & l-$95$\% & u-$95$\% \\
\midrule
\multicolumn{6}{c}{\texttt{AUROC}}\\
Linespots & 0.0166 & 0.0165 & 0.00221 & 0.0127 & 0.0215\\
Bugspots & 0.0118 & 0.0116 & 0.00156 & 0.00889 & 0.0153\\
\midrule
\multicolumn{6}{c}{\texttt{EXAM}}\\
Linespots & 0.209 & 0.209 & 0.00745 & 0.195 & 0.224\\
Bugspots & 0.225 & 0.225 & 0.00779 & 0.21 & 0.241\\
\midrule
\multicolumn{6}{c}{\texttt{AUCEC5}}\\
Linespots & 0.231 & 0.23 & 0.0148 & 0.202 & 0.261\\
Bugspots & 0.159 & 0.159 & 0.0114 & 0.137 & 0.182\\
\midrule
\multicolumn{6}{c}{\texttt{AUCEC1}}\\
Linespots & 0.102 & 0.102 & 0.00747 & 0.0879 & 0.117\\
Bugspots & 0.0627 & 0.0625 & 0.00524 & 0.0529 & 0.0735\\
\midrule
\multicolumn{6}{c}{$E_{\mr{inspect}}@100$}\\
Linespots & 0.341 & 0.335 & 0.0877 & 0.187 & 0.529\\
Bugspots & 0.213 & 0.209 & 0.0563 & 0.115 & 0.334\\
\midrule
\multicolumn{6}{c}{$E_{\mr{inspect}}@10$}\\
Linespots & 0.0505 & 0.0484 & 0.0174 & 0.0231 & 0.0907\\
Bugspots & 0.034 & 0.0324 & 0.0119 & 0.0153 & 0.0614\\
\midrule
\multicolumn{6}{c}{$E_{\mr{inspect}}F$}\\
Linespots & 310 & 301 & 79.5 & 180 & 490\\
Bugspots & 1318 & 1279 & 337 & 765 & 2097\\
\midrule
\multicolumn{6}{c}{\texttt{Runtime}}\\
Linespots & 11.3 & 11.2 & 1.73 & 8.26 & 15\\
Bugspots & 0.0115 & 0.0115 & 0.000784 & 0.0101 & 0.0132\\
\bottomrule
\end{tabular}
\end{table}

\subsection{RQ2: Runtime comparison}
The results of the runtime model need some additional information. Initially, we opted for designing relatively simple generalized linear models (GLMs). However, when we conducted posterior predictive checks, they indicated an inferior fit. The main reason was that the posterior consisted of three modes (i.e., three distinct peaks in the posterior probability distribution).

\begin{figure}
    \centering
    \subfloat[Outcome scale effect of Bugspots for both mixture components]{
        \includegraphics[width=0.45\textwidth]{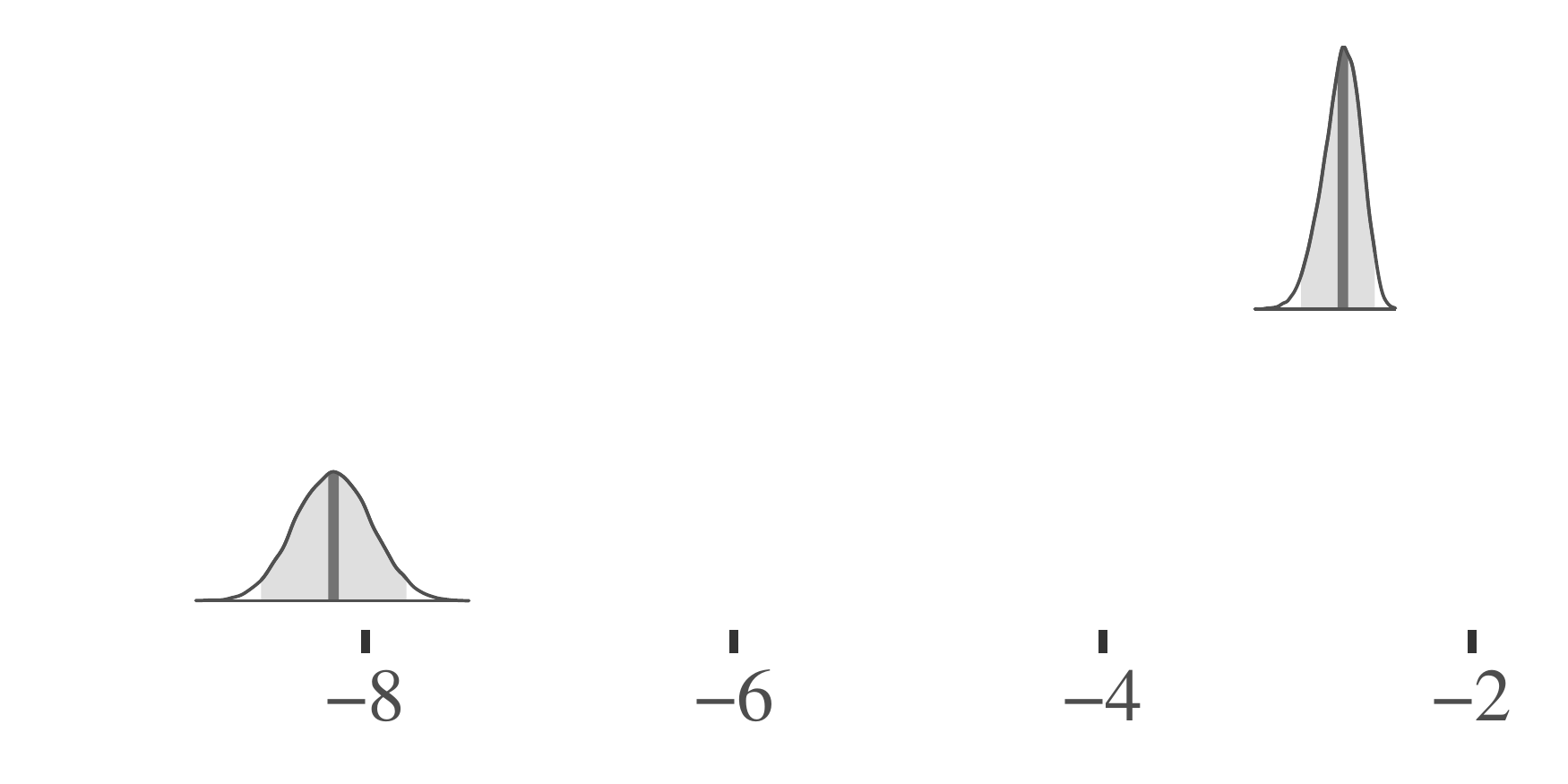}
    }
    \hfill
    \subfloat[Conditional effects]{
        \includegraphics[width=0.45\textwidth]{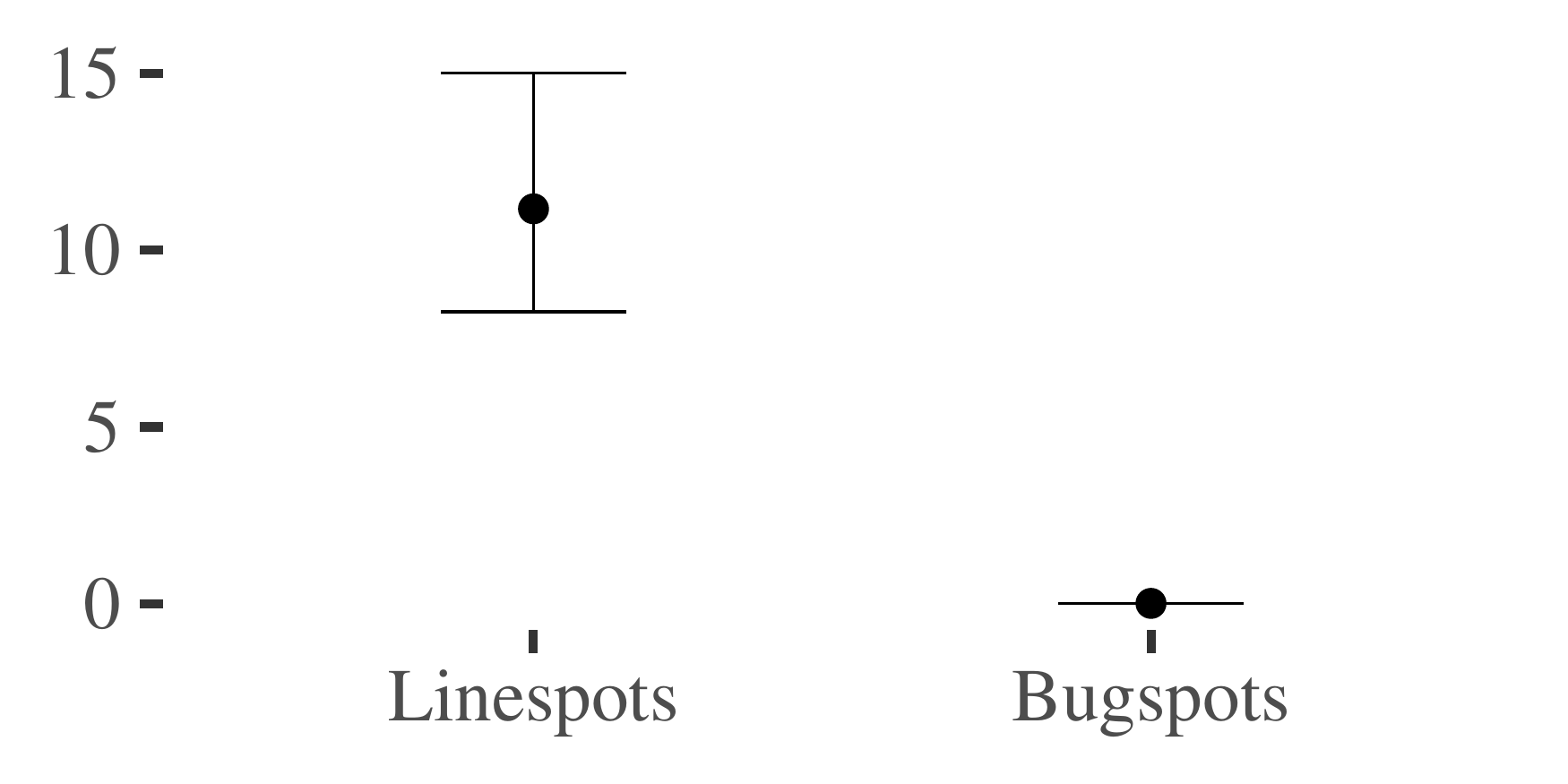}
    }
    \caption{Effects of algorithm on runtime}
    \label{fig:res-runtime}
\end{figure}

Based on the results of the simple GLMs, and our experience when developing and testing the algorithms, we assumed that two different phenomena caused the three modes.
The first two modes would be the well-behaving samples that would execute quickly for both algorithms. The third mode would be caused by samples that include big commits, likely with formatting changes or generated code across many files, that take substantially longer to generate diffs for and parse them.
Based on these assumptions, we built a mixture model for the runtime using two shifted-lognormal likelihoods.
Looking at the effects in Figure~\ref{fig:res-runtime}, both components estimate Bugspots to have reduced runtime compared to Linespots with no $0$ overlap. (The runtime effect sizes are the only ones with an identity link, meaning the effects are on the outcome scale.) The conditional effects show the same, with Linespots consistently having longer runtimes than Bugspots.
This follows our expectation, as Linespots should have longer execution time. After all, Linespots does everything Bugspots does but adds diff generation and parsing on top of it.

Table~\ref{tab:conditional-effects-summary} shows the summary of the model results concerning runtime performance. In the table, one will find a summary of the estimated conditional effects for all eight evaluation metrics for both algorithms.
We were able to show that Linespots outperforms Bugspots consistently for different ranges of the result list while taking substantially longer to compute.

\section{Threats to validity}
\label{sec:threats-to-validity}
As the results of this study are based on the confidence we have in the dataset, and especially the validation data we collected ourselves, it is necessary to ensure their accuracy.
Recall, that we added $10$ projects to our dataset that used stringent commit message guidelines and consistently enforced them. We call the group of samples from those $10$ projects `good', while we call the group of samples from projects using no CMC `base'. 

To ensure that our conclusions are not based on a subpar data collection procedure, we can compare the results of the \texttt{good} samples with those of the \texttt{base} samples. We also show the individual CMCs to show how the individual CMCs compare to the \texttt{good} and \texttt{base} groups.

\begin{figure}
    \centering
    \subfloat[]{
        \includegraphics[width=0.45\textwidth]{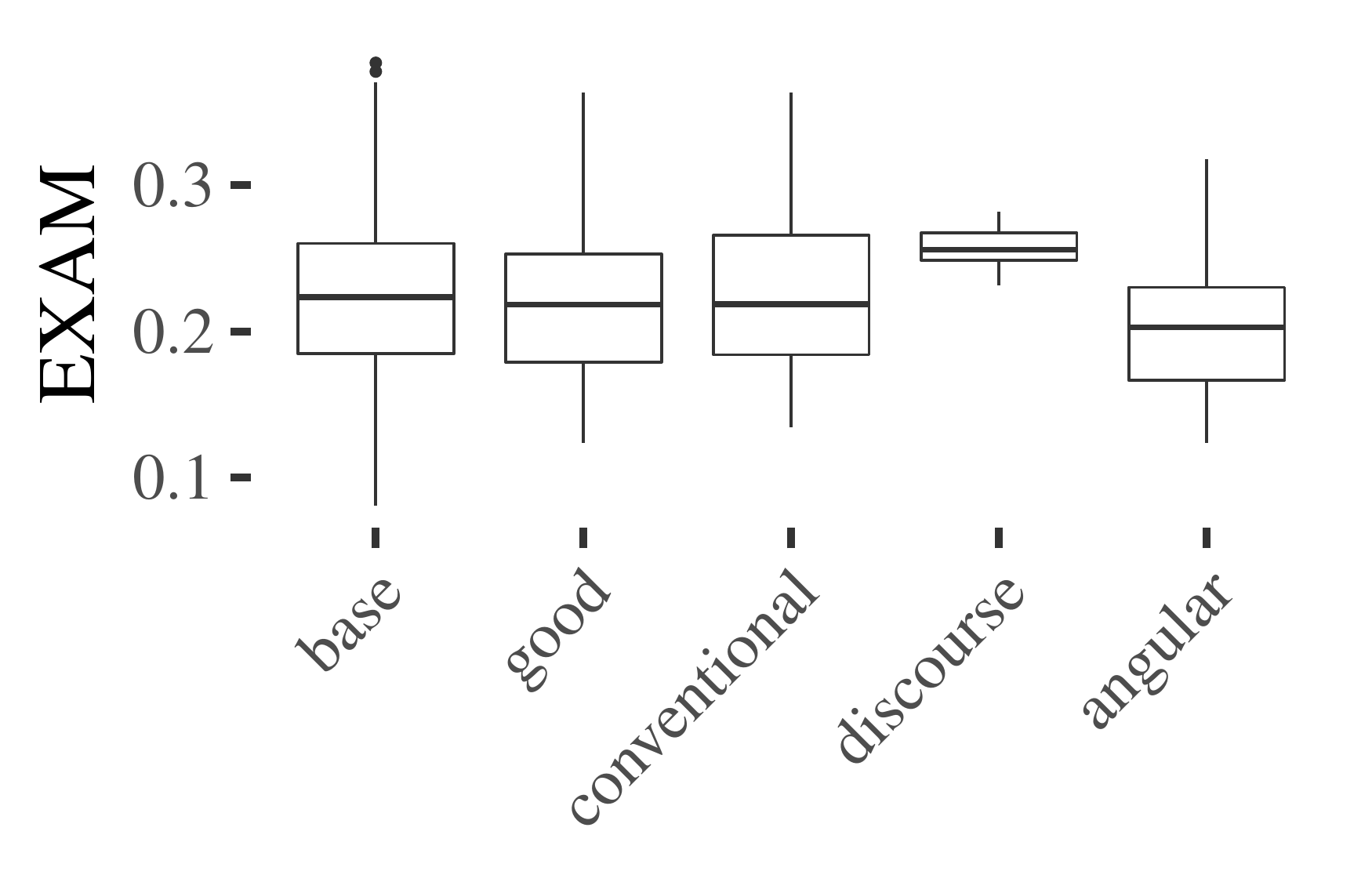}
    }
    \hfill
    \subfloat[]{
        \includegraphics[width=0.45\textwidth]{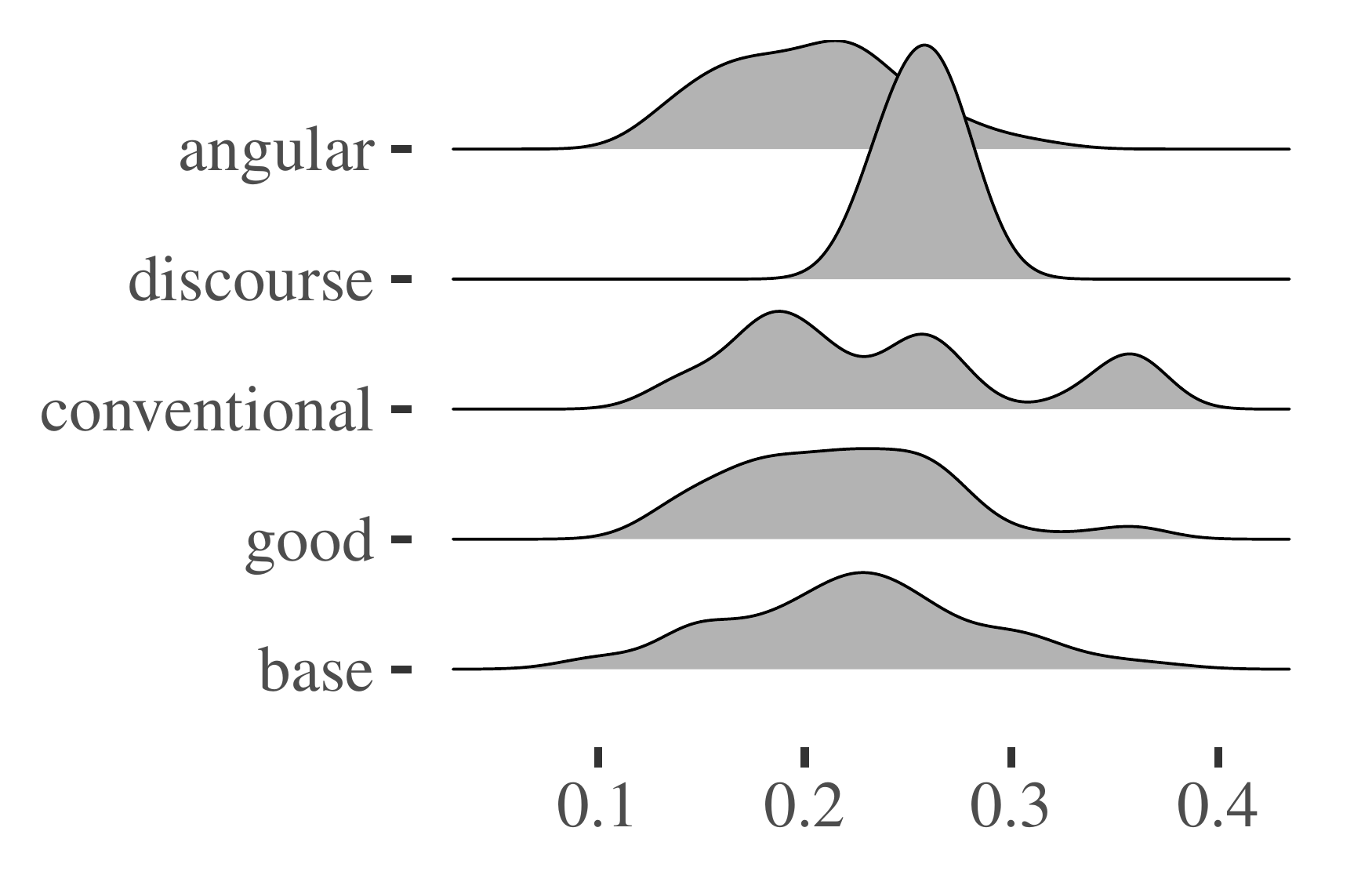}
    }
    \caption{Comparison of CMCs on \texttt{EXAM} score}
    \label{fig:threats-1}
\end{figure}

Figure~\ref{fig:threats-1} shows a simple comparison for the \texttt{EXAM} score between the two CMC groups and the three individual CMCs.
There are differences between the individual CMCs and the \texttt{base} group, especially for the discourse samples.
But when combined into the \texttt{good} group, the differences average out and the result looks very similar to the \texttt{base} group. This is probably caused by the relatively small number of samples per CMC compared to the \texttt{base} group.

The similarity between the \texttt{base} and the \texttt{good} group does support our method of data collection.
While we would expect more uncertainty in the \texttt{good} group compared to the \texttt{base} group based on the smaller sample size, $132$ compared to $348$, respectively, the higher quality could cause the reduction.
Nevertheless, we also built a model to investigate the effect of the CMC on the result, and the effect of the algorithm on the result.

\begin{figure} 
    \centering
    \subfloat[Conditional effect per CMC for Bugspots (1) and Linespots (2)]{
    \label{fig:threats-2-a}
        \includegraphics[width=0.45\textwidth]{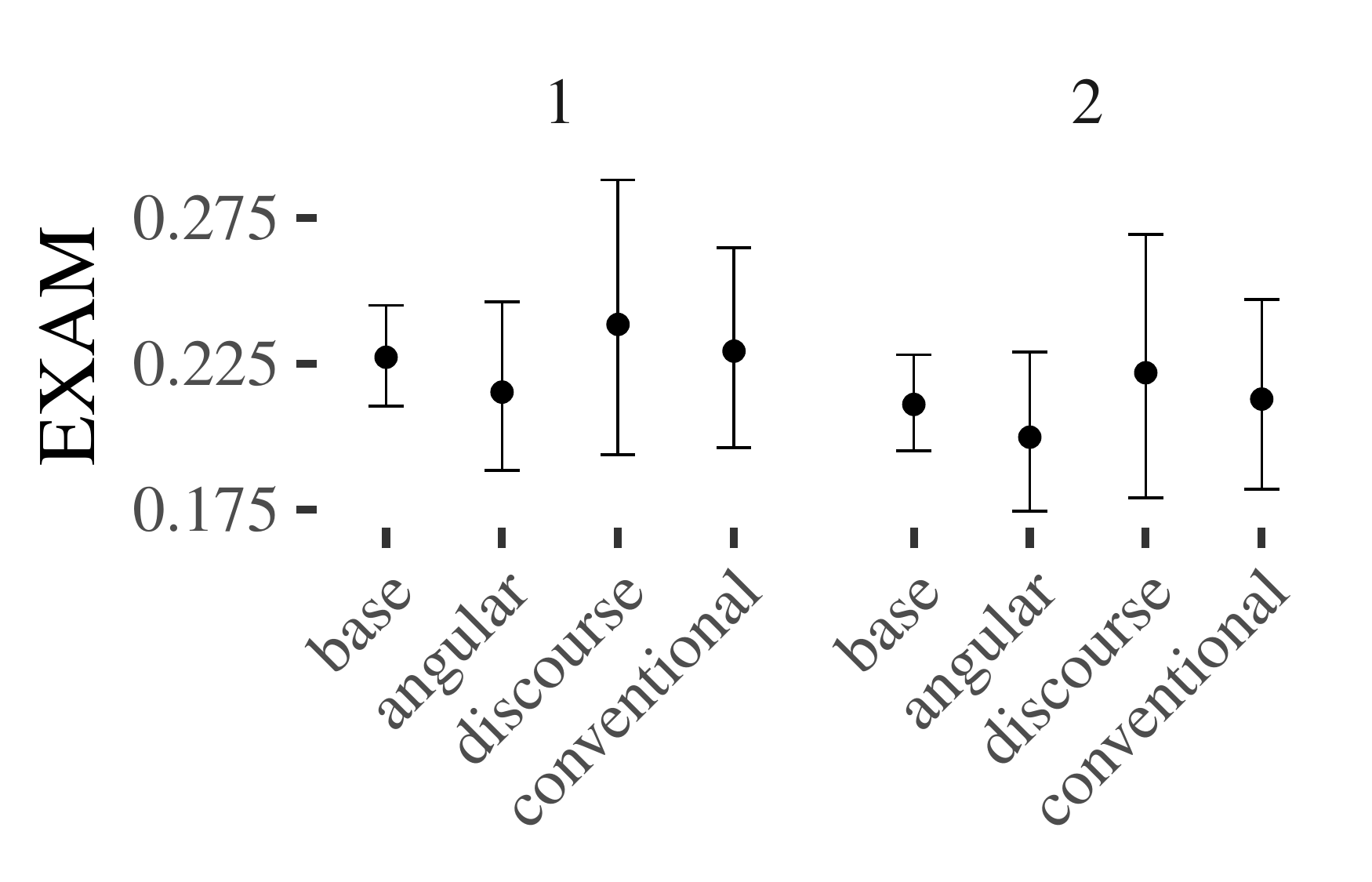}
    }
    \hfill
    \subfloat[Logit scale effect of the different CMCs]{
    \label{fig:threats-2-b}
        \includegraphics[width=0.45\textwidth]{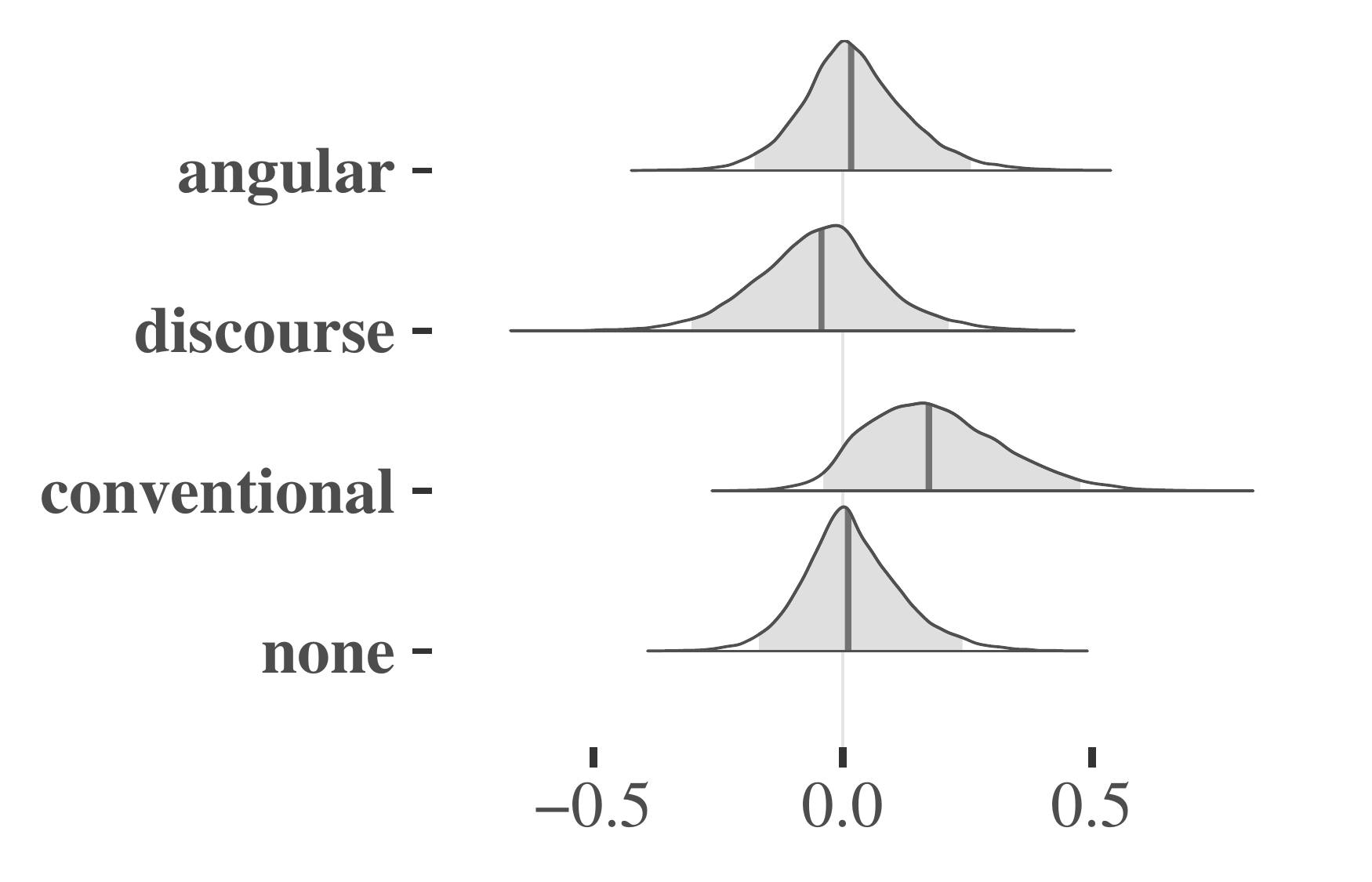}
    }
    \caption{Effects of CMCs on \texttt{EXAM}}
    \label{fig:threats-2}
\end{figure}

Figure~\ref{fig:threats-2-a} shows the conditional effects of the CMCs on the \texttt{EXAM} score for Bugspots $(1)$ and Linespots $(2)$, and while there is variation in the means and tails, the overlap between CMCs is considerable. The small sample size could simply cause the higher uncertainty in the \texttt{good} quality groups.
It is also easy to see that Linespots produces smaller \texttt{EXAM} scores for every case.
The $\logit$ scale effect of the CMCs on the algorithm effect in Figure~\ref{fig:threats-2-b} shows similar behavior. Only the conventional commits convention is close to having a significant effect.

Another possible threat is the assumptions we made when developing the DAG\@.
The two critical assumptions are that there is no incoming arrow for algorithm and only a single outgoing one, towards the evaluation metrics.
We are confident that there are no incoming arrows to algorithm, as we ran every sample with both Bugspots and Linespots, so nothing influenced which sample used which algorithm.
The second assumption would not hold if there were an unknown variable that the entity algorithm influences. We argue that the only output of the algorithms is the result list, which directly leads to all the performance-related evaluation metrics and the runtime of the algorithm itself. While there could be an interaction with the memory use of the algorithm, we assume it to be neglectable unless it would lead to out-of-memory problems.

Finally, when we checked the bad smells of software analytics by Menzies and Shepperd~\cite{menziesBadSmellsSoftware2018}, we find that all of the frequentist statistics problems are non-issues due to our use of Bayesian data analysis and reporting of posteriors and credible intervals. Furthermore, with us publishing both our entire code, data, and complete replication packages~\cite{LinespotsReferenceImplementation, scholzLinespotsReproducabilityDocker2020,scholzLinespotsAnalysisRepository2019}, we actively prevent any of the presentation related problems as listed by Menzies and Shepperd.

\section{Discussion}
\label{sec:discussion}
Starting with the performance, Linespots does perform better than Bugspots for all evaluation metrics in the average case scenario. The summary for conditional effect sizes is shown in Table~\ref{tab:conditional-effects-summary}.
Starting with  \texttt{AUROC} and \texttt{EXAM}, as both are averaging across the entire result list, there is much overlap for the two algorithms on the outcome scale. As the effects on the $\logit$ scale, shown in Figures~\ref{fig:res-auroc}--\ref{fig:res-exam}, have no $0$ overlap in both cases, the overlap on the outcome scale is caused by uncertainty propagated by the model.
Both algorithms still perform poorly, and the absolute differences are small, with a $0.0059$ increase in \texttt{AUROC} and a $0.016$ decrease in \texttt{EXAM} for Linespots. However, the relative improvements of Linespots over Bugspots are substantial, with a $40$\% increase in \texttt{AUROC} and a $7$\% decrease in \texttt{EXAM}.

Moving to \texttt{AUCEC5} and \texttt{AUCEC1}, the differences between the algorithms become even clearer as Linespots outperforms Bugspots consistently in both measures (both on the $\logit$ and outcome scale) as shown in Figures~\ref{fig:res-aucec5}--\ref{fig:res-aucec1}.
Here the trend starts to show that Linespots' relative performance to Bugspots improves, the further up the result list one measure. With an advantage of $0.0395$ or $63$\% for \texttt{AUCEC1}, $0.069$ or $45$\% for \texttt{AUCEC5} and with both effects being far from $0$ on the $\logit$ scale.

When looking at the very first entries on the result list with the $E_{\mr{inspect}}@10$ and $E_{\mr{inspect}}@100$ evaluation metrics, the picture becomes less certain again. While Linespots still outperforms Bugspots on average, with $0.128$ or $60$\% improved $E_{\mr{inspect}}@100$ and $0.0165$ or $49$\% improved $E_{\mr{inspect}}@100$ on average, there is now some zero overlap in the tails on the $\logit$ scale and wide overlap on the outcome scale as Figures~\ref{fig:res-einspect100}--\ref{fig:res-einspect10} show. We assume that this is due to both algorithms just not performing well enough to make the $E_{\mr{inspect}}@1n$ evaluation metrics reliable with such small values. There are many cases where both algorithms predict zero faults in the first $10$ or $100$ results.
However, Linespots still performs better on average than Bugspots. As the zero overlap on the $\logit$ scale is only in the tails, we still conclude that Linespots offers better $E_{\mr{inspect}}@10$ and $E_{\mr{inspect}}@100$ performance than Bugspots.

Finally, we look at $E_{\mr{inspect}}F$, which again shows a definite advantage to Linespots in Figure~\ref{fig:res-einspectf}. Bugspots has a clear negative effect on the log scale and Linespots predicts the first fault $1008$ lines before Bugspots on average, or in just $24$\% of the lines of what Bugspots needs.
While $310$ lines on average is still a lot, it moves a lot closer to the $200$ lines threshold Long and Rinard~\cite{longAnalysisSearchSpaces2016} found to be of value.

Together, all seven evaluation metrics show clear improvements for Linespots over Bugspots in predictive performance. Both, when averaged across all faults, but especially for the result list's critical early parts.

The second aspect that made Bugspots an interesting algorithm was that it was simple and easy to compute.
As Linespots do more computations than Bugspots, we expect the runtime to increase, which is reflected in the results.
Bugspots is many times faster than Linespots, even with the more complex effects from the mixture model.
Moreover, the results of Linespots are somewhat bloated by a smaller number of very slow running samples, which is a drawback of Linespots. Ultimately, we believe that the runtime of Linespots can be improved by refactoring the code with a particular emphasis on performance; however, it will likely never be as fast as Bugspots, and perhaps that is not even important.

While fast runtimes are usually desirable, there is some leeway in how long an analysis can take before it becomes a burden to developers. This heavily depends on where in the workflow, something is done.
Some things, like syntax highlighting, auto-completion, or simple linting, are usually expected to give the developer real-time feedback. A simple unit test suite could take a few seconds or minutes, while more complex tests or analyses can take a lot longer.
The usual scenario for fault prediction algorithms is to focus code review or testing efforts. In both of these cases, the algorithm would run once, and then, based on the result, the next steps will be decided.
For this kind of work, that could easily be added to a continuous integration pipeline, even the most extended runtimes of a few minutes for Linespots would be feasible.
This assessment is supported by the findings of Kochhar et al.~\cite{kochharPractitionersExpectationsAutomated2016} who showed that runtimes of under one minute are acceptable by more than $90$\% of developers.

\section{Related Work}
\label{sec:related-work}
Before presenting specific studies, we want to point out that a comparison of results between studies can be complicated. One of the main reasons is that the datasets differ between studies, and without standardized implementations, the calculation of both fault prediction and evaluation metrics could deviate.

This can range from slightly different implementations of the same algorithm to problems like differing granularities.
When Bugspots predicts a fault by proposing a file that potentially contains a fault, it is not clear how to compare that to Linespots predicting a fault by pointing to an individual line in a file.
We tried to reduce these problems by mapping Bugspots' results to lines, following Zou et al.~\cite{zouEmpiricalStudyFault2019} as discussed in Section~3.1.6.

When comparing our results to Zou et al.~\cite{zouEmpiricalStudyFault2019}, we can see that not all of the results match.
Starting with the runtime, Bugspots performed a lot faster on our test system than it did for them. However, this could simply be down to the used hardware or implementation. In both cases, the runtime is under a second.
When comparing $E_{\mr{inspect}}@10$ results, the $0.03$ average that Bugspots achieved in our study is not far from the $0$ by them. The difference can easily come from the dataset. When we ran a model with the projects both studies have in common, Bugspots did not find a single fault as well.

However, when comparing EXAM results, our values are roughly half of what Zou et al.~\cite{zouEmpiricalStudyFault2019} report. This persists when running a model with just the common projects.
We assume that this is due to the differences in validation data. While we have not tested how similar our approach is to the Defects4J set, we expect there to be more defects in our approach. While it is not clear if those are false positives due to ambiguous commit messages, or true faults that are not part of Defects4J, it could explain why our study showed better EXAM results.
If we assume that there are no significant differences between the two Bugspots implementations and use the difference between Bugspots' results to scale to our Linespots' results, Linespots would lie between Bugspots and BugLocator in terms of performance.

D'Ambros et al.~\cite{dambrosEvaluatingDefectPrediction2012} compared different fault prediction metrics, including the number of past faults metric, which Bugspots is based on.
We expect Bugspots to perform better than the NFIX-ONLY metric, due to the added weight decay.
The $p_{\mathrm{eff}}$ mean result of $0.66$, by D'Ambros et al.~\cite{dambrosEvaluatingDefectPrediction2012}, is within the range of what we see reasonable when compared to the $0.78$ result of the \texttt{AUCEC100} model (not reported due to redundancy with \texttt{EXAM} results as discussed in Section~3.1.4). While not all of the differences can be attributed to the weight decay, i.e., different projects and granularities used in the studies, it allow us to set Linespots in perspective.
The best performing fault prediction metric LDHH has a mean $p_{\mathrm{eff}}$ of $0.81$, which is close to Linespots' $0.80$ result in our study.

It is important to remember that we can not merely compare these values as they are gathered with different methods. However, the results indicate that Linespots is competitive with some of the better performing fault prediction metrics in the field and should serve as part of fault prediction models.

\section{Conclusions}
\label{sec:conclusion}
In this study, we proposed Linespots, a novel variant of the Bugspots fault prediction algorithm, and evaluated it by investigating two research questions.

RQ1: How does the predictive performance of Linespots compare to Bugspots?
We found that Linespots outperform Bugspots on all evaluation metrics regarding predictive performance, especially when focusing on the result lists' earlier parts. This is important as developers and tools will only look at the very early parts of a result list~\cite{parninAreAutomatedDebugging2011a}.

While the overall performance is not good enough to be useful in a code review scenario, in our opinion, Linespots can serve as an improved Baseline when evaluating new techniques and as an essential part of fault prediction models.
We also agree with Lewis et al.~\cite{lewisDoesBugPrediction2013} in their assessment that fault prediction is more suitable to support focusing testing efforts instead of code reviews. This will be true until fault prediction can offer actionable results to developers, in our opinion.

RQ2: How does the runtime performance of Linespots compare to Bugspots?
We found that Linespots can take a lot longer to run than Bugspots, as was somewhat expected. While the runtime might be too long in some extreme cases, we assume that projects with that kind of size will already have substantial testing suites and continuous integration pipelines, so even a runtime of a few minutes might not affect the possibility to include Linespots in their setup. However, it might be interesting to investigate what exactly causes the big spikes in runtime and how much the runtime can be reduced.
When looking at the comparison by Zou et al.~\cite{zouEmpiricalStudyFault2019}, Linespots would still be substantially faster than most fault prediction methods, which is why we conclude that it does fill a similar role to Bugspots in regards to runtime.

Based on the related work comparison and our findings, we argue that Linespots is a well-performing fault prediction metric. Combined with the performance improvements over Bugspots, we recommend using Linespots over Bugspots in all cases where real-time performance is not needed.
We can confidently attribute the performance improvements to the change in granularity between Bugspots and Linespots. The only other differences between the two algorithms are some additional corner-case handling for Linespots that are not necessary for Bugspots.

\section*{Data Availability Statement }
The data that support the findings of this study are openly available in the linespots-analysis repository at \url{https://github.com/sims1253/linespots-analysis/} together with all code and a replication package.

\ack
We want to thank the Stan (\url{https://discourse.mc-stan.org}) community for their valuable feedback and support during the analysis.

The computations were enabled by resources provided by the Swedish National Infrastructure for Computing (SNIC), partially funded by the Swedish Research Council through grant agreement no.\ 2018--05973.

\bibliographystyle{wileyj}
\bibliography{references}

\begin{thebibliography}{10}
\providecommand{\url}[1]{\texttt{#1}}
\providecommand{\urlprefix}{URL }
\expandafter\ifx\csname urlstyle\endcsname\relax
  \providecommand{\doi}[1]{doi:\discretionary{}{}{}#1}\else
  \providecommand{\doi}{doi:\discretionary{}{}{}\begingroup
  \urlstyle{rm}\Url}\fi

\bibitem{hallSystematicLiteratureReview2012}
Hall T, Beecham S, Bowes D, Gray D, Counsell S. A systematic literature review
  on fault prediction performance in software engineering. \emph{IEEE
  Transactions on Software Engineering}  Nov 2012; \textbf{38}(6):1276--1304,
  \doi{10.1109/TSE.2011.103}.

\bibitem{radjenovicSoftwareFaultPrediction2013}
Radjenovi{\'c} D, Heri{\v c}ko M, Torkar R, {\v Z}ivkovi{\v c} A. Software
  fault prediction metrics: {{A}} systematic literature review.
  \emph{Information and Software Technology}  2013; \textbf{55}(8):1397--1418,
  \doi{10.1016/j.infsof.2013.02.009}.

\bibitem{hassanTopTenList2005}
Hassan AE, Holt RC. The top ten list: Dynamic fault prediction. \emph{21st
  {{IEEE International Conference}} on {{Software Maintenance}} ({{ICSM}}'05)},
  2005; 263--272, \doi{10.1109/ICSM.2005.91}.

\bibitem{kimPredictingFaultsCached2007a}
Kim S, Zimmermann T, Whitehead~Jr EJ, Zeller A. Predicting faults from cached
  history. \emph{29th {{International Conference}} on {{Software Engineering}}
  ({{ICSE}}'07)}, 2007; 489--498, \doi{10.1109/ICSE.2007.66}.

\bibitem{rahmanBugCacheInspectionsHit2011}
Rahman F, Posnett D, Hindle A, Barr E, Devanbu P. {BugCache} for inspections:
  {H}it or miss? \emph{19th {{ACM SIGSOFT Symposium}} and 13th {{European
  Conference}} on {{Foundations}} of {{Software Engineering}}},
  {{ESEC}}/{{FSE}} '11, {ACM}: {New York, NY, USA}, 2011; 322--331,
  \doi{10.1145/2025113.2025157}.

\bibitem{dambrosExtensiveComparisonBug2010}
D'Ambros M, Lanza M, Robbes R. An extensive comparison of bug prediction
  approaches. \emph{2010 7th {{IEEE Working Conference}} on {{Mining Software
  Repositories}} ({{MSR}} 2010)}, {IEEE}: {Cape Town, South Africa}, 2010;
  31--41, \doi{10.1109/MSR.2010.5463279}.

\bibitem{dambrosEvaluatingDefectPrediction2012}
D'Ambros M, Lanza M, Robbes R. Evaluating defect prediction approaches: A
  benchmark and an extensive comparison. \emph{Empirical Software Engineering}
  Aug 2012; \textbf{17}(4-5):531--577, \doi{10.1007/s10664-011-9173-9}.

\bibitem{lewisDoesBugPrediction2013}
Lewis C, Lin Z, Sadowski C, Zhu X, Ou R, Whitehead EJ. Does bug prediction
  support human developers? {{Findings}} from a {{Google}} case study.
  \emph{35th {{International Conference}} on {{Software Engineering}}
  ({{ICSE}})}, 2013; 372--381, \doi{10.1109/ICSE.2013.6606583}.

\bibitem{kochharPractitionersExpectationsAutomated2016}
Kochhar PS, Xia X, Lo D, Li S. Practitioners' expectations on automated fault
  localization. \emph{Proceedings of the 25th {{International Symposium}} on
  {{Software Testing}} and {{Analysis}}}, {{ISSTA}} 2016, {Association for
  Computing Machinery}: {Saarbr\"ucken, Germany}, 2016; 165--176,
  \doi{10.1145/2931037.2931051}.

\bibitem{hataBugPredictionBased2012}
Hata H, Mizuno O, Kikuno T. Bug prediction based on fine-grained module
  histories. \emph{34th {{International Conference}} on {{Software
  Engineering}} ({{ICSE}})}, 2012; 200--210, \doi{10.1109/ICSE.2012.6227193}.

\bibitem{shirabadPROMISERepositorySoftware2005}
Shirabad JS, Menzies TJ. The {{PROMISE}} repository of software engineering
  databases. \emph{School of Information Technology and Engineering, University
  of Ottawa, Canada}  2005; \textbf{24}.

\bibitem{justDefects4JDatabaseExisting2014}
Just R, Jalali D, Ernst MD. {{Defects4J}}: {A} database of existing faults to
  enable controlled testing studies for {{Java}} programs. \emph{2014
  {{International Symposium}} on {{Software Testing}} and {{Analysis}}},
  {{ISSTA}} 2014, {Association for Computing Machinery}: {San Jose, CA, USA},
  2014; 437--440, \doi{10.1145/2610384.2628055}.

\bibitem{wassersteinMovingWorld052019}
Wasserstein RL, Schirm AL, Lazar NA. Moving to a world beyond ``$p<0.05$''.
  \emph{The American Statistician}  Mar 2019; \textbf{73}(sup1):1--19,
  \doi{10.1080/00031305.2019.1583913}.

\bibitem{liEvaluatingSoftwareDefect2019}
Li L, Lessmann S, Baesens B. Evaluating software defect prediction performance:
  {A}n updated benchmarking study. \emph{SSRN Electronic Journal}  2019;
  \doi{10.2139/ssrn.3312070}.

\bibitem{LinespotsReferenceImplementation}
Linespots {{Reference Implementation}}.
  https://gitlab.com/sims1253/linespots-lib.

\bibitem{scholzLinespotsReproducabilityDocker2020}
Scholz M. Replication package. https://github.com/sims1253/linespots-docker.

\bibitem{scholzLinespotsAnalysisRepository2019}
Scholz M. Analysis repository. https://github.com/sims1253/linespots-analysis.

\bibitem{wangVersionHistorySimilar2014}
Wang S, Lo D. Version history, similar report, and structure: Putting them
  together for improved bug localization. \emph{22nd {{International
  Conference}} on {{Program Comprehension}}}, {{ICPC}} 2014, {Association for
  Computing Machinery}: {Hyderabad, India}, 2014; 53--63,
  \doi{10.1145/2597008.2597148}.

\bibitem{youmImprovedBugLocalization2017}
Youm KC, Ahn J, Lee E. Improved bug localization based on code change histories
  and bug reports. \emph{Information and Software Technology}  Feb 2017;
  \textbf{82}:177--192, \doi{10.1016/j.infsof.2016.11.002}.

\bibitem{zouEmpiricalStudyFault2019}
Zou D, Liang J, Xiong Y, Ernst MD, Zhang L. An empirical study of fault
  localization families and their combinations. \emph{IEEE Transactions on
  Software Engineering}  2019; :1--1\doi{10.1109/TSE.2019.2892102}.

\bibitem{grigorikImplementationSimpleBug2019}
Grigorik I. Implementation of simple bug prediction hotspot heuristic:
  Igrigorik/bugspots Aug 2019.

\bibitem{myersAnONDDifference1986}
Myers EW. {{AnO}}({{ND}}) difference algorithm and its variations.
  \emph{Algorithmica}  Nov 1986; \textbf{1}(1-4):251--266,
  \doi{10.1007/BF01840446}.

\bibitem{nugrohoHowDifferentAre2020}
Nugroho YS, Hata H, Matsumoto K. How different are different diff algorithms in
  {{Git}}? \emph{Empirical Software Engineering}  Jan 2020;
  \textbf{25}(1):790--823, \doi{10.1007/s10664-019-09772-z}.

\bibitem{fawcettROCGraphsNotes2004}
Fawcett T. {ROC} graphs: {N}otes and practical considerations for data mining
  researchers. \emph{ReCALL}  Jan 2004; \textbf{31}:1--38.

\bibitem{arisholmDataMiningTechniques2007}
Arisholm E, Briand LC, Fuglerud M. Data mining techniques for building
  fault-proneness models in telecom {J}ava software. \emph{18th {{IEEE
  International Symposium}} on {{Software Reliability}} ({{ISSRE}} '07)}, 2007;
  215--224, \doi{10.1109/ISSRE.2007.22}.

\bibitem{arisholmSystematicComprehensiveInvestigation2010}
Arisholm E, Briand LC, Johannessen EB. A systematic and comprehensive
  investigation of methods to build and evaluate fault prediction models.
  \emph{Journal of Systems and Software}  Jan 2010; \textbf{83}(1):2--17,
  \doi{10.1016/j.jss.2009.06.055}.

\bibitem{parninAreAutomatedDebugging2011a}
Parnin C, Orso A. Are automated debugging techniques actually helping
  programmers? \emph{{{International Symposium}} on {{Software Testing}} and
  {{Analysis}} {{(ISSTA)}}}, {ACM Press}: {Toronto, Ontario, Canada}, 2011;
  199, \doi{10.1145/2001420.2001445}.

\bibitem{longAnalysisSearchSpaces2016}
Long F, Rinard M. An analysis of the search spaces for generate and validate
  patch generation systems. \emph{38th {{International Conference}} on
  {{Software Engineering}} ({{ICSE}})}, 2016; 702--713,
  \doi{10.1145/2884781.2884872}.

\bibitem{b.leLearningtorankBasedFault2016a}
B~Le TD, Lo D, Le~Goues C, Grunske L. A learning-to-rank based fault
  localization approach using likely invariants. \emph{25th {{International
  Symposium}} on {{Software Testing}} and {{Analysis}} ({ISSTA})}, {ACM Press}:
  {Saarbr\&\#252;cken, Germany}, 2016; 177--188, \doi{10.1145/2931037.2931049}.

\bibitem{wongCrosstabbasedStatisticalMethod2008}
Wong E, Wei T, Qi Y, Zhao L. A crosstab-based statistical method for effective
  fault localization. \emph{1st {{International Conference}} on {{Software
  Testing}}, {{Verification}}, and {{Validation}}}, 2008; 42--51,
  \doi{10.1109/ICST.2008.65}.

\bibitem{baltesSamplingSoftwareEngineering2020}
Baltes S, Ralph P. Sampling in software engineering research: {A} critical
  review and guidelines. \emph{arXiv:2002.07764 [cs]}  Feb 2020; .

\bibitem{tothPublicBugDatabase2016}
T{\'o}th Z, Gyimesi P, Ferenc R. A public bug database of {GitHub} projects and
  its application in bug prediction. \emph{Computational {{Science}} and {{Its
  Applications}} -- {{ICCSA}}}, vol. 9789, Gervasi O, Murgante B, Misra S,
  Rocha AMA, Torre CM, Taniar D, Apduhan BO, Stankova E, Wang S (eds.).
  {Springer International Publishing}: {Cham}, 2016; 625--638,
  \doi{10.1007/978-3-319-42089-9_44}.

\bibitem{pearsonEvaluatingImprovingFault2016}
Pearson S, Campos J, Just R, Fraser G, Abreu R, Ernst MD, Pang D, Keller B.
  Evaluating \& improving fault localization techniques. \emph{University of
  Washington Department of Computer Science and Engineering, Seattle, WA, USA,
  Tech. Rep. UW-CSE-16-08-03}  2016; :27.

\bibitem{mengSystematicEditingGenerating2011}
Meng N, Kim M, McKinley KS. Systematic editing: Generating program
  transformations from an example. \emph{ACM SIGPLAN Notices}  2011;
  \textbf{46}(6):14.

\bibitem{rahmanComparingStaticBug2014}
Rahman F, Khatri S, Barr ET, Devanbu P. Comparing static bug finders and
  statistical prediction. \emph{36th {{International Conference}} on {{Software
  Engineering}} - {{ICSE}} 2014}, {ACM Press}: {Hyderabad, India}, 2014;
  424--434, \doi{10.1145/2568225.2568269}.

\bibitem{schadPrincipledBayesianWorkflow2020}
Schad DJ, Betancourt M, Vasishth S. Toward a principled {B}ayesian workflow in
  cognitive science. \emph{Psychological Methods}  2020;
  \doi{10.1037/met0000275}.

\bibitem{pearlSevenToolsCausal2019}
Pearl J. The seven tools of causal inference, with reflections on machine
  learning. \emph{Communications of the ACM}  Feb 2019; \textbf{62}(3):54--60,
  \doi{10.1145/3241036}.

\bibitem{pearlCausalityModelsReasoning2009}
Pearl J. \emph{Causality: {M}odels, reasoning and inference}. Second edn.,
  {Cambridge University Press}: {New York, NY, USA}, 2009.

\bibitem{mcelreathStatisticalRethinkingBayesian2020}
McElreath R. \emph{Statistical rethinking: {A} {B}ayesian course with examples
  in {R} and {S}tan}. {CRC Press}, 2020.

\bibitem{burknerBrmsPackageBayesian2017a}
B{\"u}rkner PC. Brms: {A}n {R} package for {B}ayesian multilevel models using
  {S}tan. \emph{Journal of Statistical Software}  Aug 2017;
  \textbf{80}(1):1--28, \doi{10.18637/jss.v080.i01}.

\bibitem{carpenterStanProbabilisticProgramming2017}
Carpenter B, Gelman A, Hoffman MD, Lee D, Goodrich B, Betancourt M, Brubaker M,
  Guo J, Li P, Riddell A. Stan: {A} probabilistic programming language.
  \emph{Journal of Statistical Software}  Jan 2017; \textbf{76}(1),
  \doi{10.18637/jss.v076.i01}.

\bibitem{vehtariLooEfficientLeaveOneOut2019}
Vehtari A, Gelman A, Gabry J. Practical {B}ayesian model evaluation using
  leave-one-out cross-validation and {WAIC}. \emph{Statistics and Computing}
  2017; \textbf{27}:1413--1432, \doi{10.1007/s11222-016-9696-4}.

\bibitem{menziesBadSmellsSoftware2018}
Menzies T, Shepperd M. ``bad smells'' in software analytics papers.
  \emph{Information and Software Technology}  2019; \textbf{112}:35--47,
  \doi{10.1016/j.infsof.2019.04.005}.

\end{thebibliography}

\end{document}